\documentclass[aip,graphicx,jcp]{revtex4-1}
\usepackage{epsfig}
\usepackage{graphics,psfrag}
\usepackage{color}
\usepackage{ulem}
\usepackage{amsmath}
\usepackage{amsfonts}
\usepackage{amssymb}

\begin{document}

\title{TWO-YUKAWA FLUID AT A HARD WALL: FIELD
THEORY TREATMENT}

\author{I. Kravtsiv}
\affiliation{Institute for Condensed Matter Physics,
National Academy of Sciences, 1 Svientsitskii Str., 79011 Lviv, Ukraine\\}
\author{T. Patsahan}
\affiliation{Institute for Condensed Matter Physics,
National Academy of Sciences, 1 Svientsitskii Str., 79011 Lviv, Ukraine\\}
\author{M. Holovko}
\affiliation{Institute for Condensed Matter Physics,
National Academy of Sciences, 1 Svientsitskii Str., 79011 Lviv, Ukraine\\}
\author{D. di Caprio}
\affiliation{Institute of Research of Chimie Paris, CNRS - Chimie ParisTech,\\
11, rue P. et M. Curie, 75005 Paris, France\\}

\date{\today}

\begin{abstract}
We apply a field-theoretical approach to study the structure and thermodynamics of a two-Yukawa fluid confined by a hard wall. We derive mean field
equations allowing for numerical
evaluation of the density profile which is compared to analytical estimations. Beyond
the mean field approximation analytical expressions for the free energy, the pressure and the
correlation function are derived. Subsequently contributions to the density profile and the adsorption coefficient due
to Gaussian fluctuations are found. Both the mean field and the fluctuation terms
of the density profile are shown to satisfy the contact theorem. We further use the contact theorem to improve the Gaussian approximation for the density profile based on a better approximation for the bulk pressure. The results obtained are compared to computer simulations data.
\end{abstract}


\maketitle
\section{Introduction}

Model systems with Yukawa-like potentials of interaction have
been extensively used for the description of a large variety of
liquids and soft matter materials. Any finite range interaction
potential between point particles can be decomposed to a sum of
Yukawa potentials with arbitrary accuracy. For instance, the Lennard-Jones potential used in the theory of simple fluids can be well approximated
by the hard repulsion with two Yukawa tails \cite{yukal1996,tang}. A hard core two-Yukawa model has been successfully used for the description
of stability of charged colloidal dispersions
\cite{wu-gao} and the properties of solutions of globular charged proteins \cite{lin}. In this case the first Yukawa term describes the
screened electrostatic interparticle repulsion and the second term approximates the Van der Waals interparticle attraction. Since the
electrostatic intercolloidal repulsion is usually more long-ranged compared to the Van der Waals attraction, such a fluid demonstrates a very rich
non-trivial phase behavior. Examples include various inhomogeneous structures such as spherical and cylindrical liquid-like clusters,
single- and multi-liquid-like slabs, cylindrical and spherical bubbles \cite{archer,archer_pini}. A hard core two-Yukawa model
was also used to explain the formation of the extra low wave vector peak in the structure factor of cytochrome C protein solutions
at moderate concentrations \cite{lin_chen}. A hard core two-Yukawa model with short-range strongly attractive interaction
was used for the description of different clusterization phenomena in associated fluids \cite{kalyuzhnyi2}. Finally, a model with
isotropic Yukawa repulsion and anisotropic Yukawa attraction has been used in the theory of nematogenic fluids \cite{holovko1999,kravtsiv2013}.
The simplicity of the Yukawa
potential allows for a description of thermodynamics and
structure of the Yukawa fluid. For hard spheres interacting
with a Yukawa tail, analytical solutions exist in the mean
spherical approximation \cite{Waisman,ginosa}. These
analytical results were generalized for the description of
hard sphere multi-Yukawa fluids \cite{hoye,lin-li}.

A model fluid of point particles with two or more Yukawa potentials is a good candidate for investigation of a fluid with an attractive interaction and soft repulsion at small distances. Such fluids with a soft repulsion have recently attracted attention particularly due to investigation of star polymers for the case when the core size of a star is small enough compared to the length of chains and the effective interaction between two stars immersed in a good solvent shows logarithmic dependence of their center-to-center separation for small distances and crosses ove
r to Yukawa form for larger ones \cite{likos,camargo}. Since Yukawa interaction is of Coulomb nature at small distances, a fluid of point particles with two Yukawa potentials can be considered as a fluid with softness intermediate between that of star polymers and simple fluids.

Yukawa models have lately been used to investigate the structure and adsorption of fluids near solid surfaces. For this aim the collective variables approach \cite{soviak}, the density field theory \cite{molphys}, the inhomogeneous integral equations approach \cite{olivares}, and the density functional theory \cite{you,tang-wu,yu,kim_kim} have been adopted. Notably in \cite{tang-wu,yu} the properties of inhomogeneous hard core two-Yukawa fluids were investigated and in \cite{kim_kim} the structure and phase behavior of the hard core model with a two-Yukawa tail potential in planar slit pores were studied.

The results for inhomogeneous fluids should satisfy certain known exact relationships, the so-called contact theorems \cite{HendersonBlumLebowitz1979,HolovkoBadialidiCaprio2005}. For a neutral fluid it states that the contact value of the point particle density near a hard wall is determined by the pressure of the fluid in the bulk volume. For an ionic fluid near a charged hard wall there is an additional electrostatic Maxwell tensor contribution. We should mention the principal difference between a fluid with the Yukawa interaction and an ionic fluid owing to the electroneutrality condition of the latter. This condition excludes some terms associated with the mean field treatment in the case of ionic fluids. In \cite{molphys} it was shown that the mean
field treatment of a Yukawa fluid near the wall reduces to
solving a non-linear differential equation for the density
profile. Different simple analytical expressions for the
density profile were obtained and compared with the numerical
estimation of the mean field results. Beyond the mean field
approximation it was shown that fluctuations can contribute
significantly to the properties of a fluid. Notably they lead
to the desorption phenomenon regardless of the sign of
interaction.

We note that the results obtained in \cite{molphys} for attractive potentials are not well defined for lower temperatures and higher densities. This problem is connected
with the divergence of the bulk correlation function along the
spinodal lines inside phase transitions of the mean field
result. Such a divergence is the result of an incorrect
treatment of short-range correlations in the bulk and can be
removed by including repulsive interactions (see for example
\cite{wheeler}). In this work we extend our previous results for the field theoretical description of a Yukawa fluid near a hard wall \cite{molphys} to the case of a fluid with two Yukawa potentials corresponding to attractive and repulsive interactions respectively.
Similar to \cite{molphys} the contributions from the mean field and from fluctuations are separated. It is shown that the mean field treatment reduces to solving a non-linear differential equation for the density profile while the treatment of Gaussian fluctuations reduces to solving the Ornstein-Zernike (OZ) integral equation with the Riemann boundary conditions. The validity of the contact theorem is verified for both contributions. However, for the bulk case the considered treatment of fluctuations leads to incorrect behavior of the pair distribution function at small interparticle distances. This also leads to overestimation of the role of fluctuations for the adsorption as well as to incorrect description of the profile near the wall. In order to improve the pair distribution function in the bulk we use the exponential approximation which gives the correct result at small distances and coincides with the previous results for larger distances. This approximation is used to calculate the bulk pressure and to improve the behavior of the density profile at small distances in the framework of the contact theorem. The quality of the obtained results is controlled by comparison with computer simulations data.

The results presented in this paper are obtained for a fluid of point particles. However, in the future we hope to modify them to describe non-point particles using the mean spherical approximation results \cite{hoye,lin-li} in a similar way as was done for non-point ionic systems \cite{holovko-soft}.

\section{The model and field theory formalism} We consider a neutral fluid of point particles
 in contact with a hard surface. The
particles do not interact with the surface but interact with
each other via a two-Yukawa potential
\begin{align}
\label{pot1}
\nu(r_{12})=\frac{A_1}{r_{12}}\,\exp(-\alpha_1r_{12})+\frac{A_2}{r_{12}}\,\exp(-\alpha_2r_{12}),
\end{align}
where $r_{12}$ denotes the distance between particles 1 and 2,
$A_1$, $A_2$ are the amplitudes of interaction and $\alpha_1$,
$\alpha_2$ are the inverse ranges. We associate the first term
of the potential with the repulsion of particles (i.e. $A_1>0$)
and the second term with the attraction ($A_2<0$). At small distances $\nu(r)=(A_1-A_2)/r_{12}>0$. As a consequence, we should have $A_1>A_2$.

In the formalism of statistical field theory the Hamiltonian
$H[\rho(\mathbf{r})]$ is a functional of field and consists of
the ideal entropy and the interaction:
\begin{align}
\label{eq1}
\beta H[\rho(\mathbf{r}_1)]&=\beta
    H^{entr}[\rho(\mathbf{r}_1)]+\beta
    H^{int}[\rho(\mathbf{r}_1)]=\\
    &\int \rho(\mathbf{r_1})\left(\ln\left[\rho(\mathbf{r_1})\Lambda^3\right]
-1\right)d\mathbf{r_1}+\nonumber\\
&\frac{\beta}{2}\int\nu(r_{12})\bigg[\rho(\mathbf{r_1})
\rho(\mathbf{r_2})-\rho(\mathbf{r}_1)\delta(\mathbf{r}_1-\mathbf{r}_2)\bigg]d\mathbf{r_1}d\mathbf{r_2},\nonumber
    \end{align}
where $\beta=1/kT$ is the inverse temperature,
$\rho(\mathbf{r})$ is the  particle density, and $\Lambda$ is
the thermal de Broglie wavelength of the particles.

As in previous papers \cite{molphys,di2003,di1998}, we adopt the
canonical ensemble approach. We fix the number of particles by
the conditions $\int \rho(\mathbf{r})d\mathbf{r}=N$ or
$\frac{1}{V}\int \rho(\mathbf{r})d\mathbf{r}=\rho_b$, where $V$
is the volume and $\rho_b=N/V$ is the average value of the bulk
density of the system. To verify this condition in a formally
unconstrained calculus we introduce a Lagrange multiplier
$\lambda$ such that
\begin{equation}
\label{MFA1}
\frac{\delta\beta
H[\rho(\mathbf{r})]}{\delta\rho(\mathbf{r})}=\lambda.
\end{equation}
The partition function $Z_N\left[\rho(\mathbf{r})\right]$ can
be expressed as
\begin{align}
\label{partition}
Z_N\left[\rho(\mathbf{r})\right]=\int\textit{D}\rho(\mathbf{r})\exp\{-\beta
H[\rho(\mathbf{r})]\},
\end{align}
where $\textit{D}\rho(\mathbf{r})$ denotes functional
integration over all possible density distributions such that
the total number of particles is $N$. The logarithm of the
partition function gives the Helmholtz free energy
\begin{align}
\beta F=-\ln Z_N.
\end{align}

\section{Mean field approximation}
The lowest
order approximation for the partition function is the saddle
point for the functional integral (\ref{partition}) which leads
to the mean field approximation (MFA).

The condition for the mean field approximation is
\begin{align}
\label{linear}
 \frac{\delta\beta H}{\delta\rho}\bigg\vert_{\rho^{MFA}(\mathbf{r})}=\lambda.
 \end{align}
In our case equation (\ref{linear}) reads
\begin{align}
\label{MFd}
\ln\frac{\rho(\mathbf{r_1})}{\rho_b}+V_1(\mathbf{r_1})+V_2(\mathbf{r_1})=\lambda,
\end{align}
where potentials $V_i(\mathbf{r_1})$ are defined as
\begin{align}
\label{f1}
V_i(\mathbf{r_1})&=\beta\int\rho(\mathbf{r_2})\frac{A_i}{r_{12}}\,
\exp(-\alpha_i r_{12})d\mathbf{r_2},\qquad i=1,2.
\end{align}
We put
\begin{align}
\lambda\equiv V_{1b}+V_{2b},
\end{align}
where $V_{ib}$ are the values of potentials $V_i(\mathbf{r}_1)$
in the bulk:
\begin{align}
\label{f2}
V_{1b}=\frac{\varkappa_1^2}{\alpha_1^2};\qquad V_{2b}=\frac{\varkappa_2^2}{\alpha_2^2},
\end{align}
and $\varkappa_i^2\equiv 4\pi\rho_b\beta A_i$.

 The gradient of
(\ref{MFd}) gives
\begin{align}
\label{grad}
\frac{\nabla\rho(\mathbf{r})}{\rho(\mathbf{r})}-\mathbf{E}_1(\mathbf{r})-\mathbf{E}_2(\mathbf{r})=0,
\end{align}
where we define an equivalent of the electric field by
\begin{align}
\mathbf{E_1}({\mathbf{r_1}})\equiv -\nabla V_1(\mathbf{r_1});\qquad
\mathbf{E_2}({\mathbf{r_1}})\equiv -\nabla V_2(\mathbf{r_1}).
\end{align}
Due to the properties of Yukawa potential
\begin{align}
\label{hered}
\left(\triangle-\alpha_1^2\right)V_1(\mathbf{r})&=-4\pi\beta A_1 \rho(\mathbf{r});\\
\label{hered1}
\left(\triangle-\alpha_2^2\right)V_2(\mathbf{r})&=-4\pi\beta A_2\rho(\mathbf{r}).
\end{align}
Replacing (\ref{hered}) and (\ref{hered1}) into (\ref{grad})
and using translational invariance parallel to the wall we
obtain
\begin{align}
\label{invariant1}
\frac{\textrm{d}}{\textrm{dz}}\left[\frac{\rho(z)}{\rho_b}+\frac{\alpha_1^2}{2\varkappa_1^2}\,
\left[V_1(z)\right]^2-\frac{1}{2\varkappa_1^2}E_1^2(z)+\frac{\alpha_2^2}{2\varkappa_2^2}\,
\left[V_2(z)\right]^2-\frac{1}{2\varkappa_2^2}E_2^2(z)\right]=0,
\end{align}
where $z$ is the distance between the particle and the wall.

\subsection{Contact theorem}\label{contact_mf}
In the bulk $\rho(z)\rightarrow\rho_b$, $E_i(z)\rightarrow 0$,
$V_i(z)\rightarrow V_{ib}$. From eq.~(\ref{invariant1}) we see
that the quantity in brackets is constant and therefore it can
be evaluated for instance in the bulk as
\begin{align}
1+\frac{\varkappa_1^2}{2\alpha_1^2}+\frac{\varkappa_2^2}{2\alpha_2^2}.
\end{align}
This quantity is the reduced pressure $\beta P/\rho_b$ within
MFA:
\begin{align}
\label{contactd}
\beta P = \rho_b\left(1+\frac{\varkappa_1^2}{2\alpha_1^2}+\frac{\varkappa_2^2}{2\alpha_2^2}\right).
\end{align}
Expression (\ref{contactd}) is the mean field approximation
which corresponds to the Van der Waals contribution. Outside
the system, where there are no particles, we have another
invariant which is simply ${\alpha_1^2}\,
V_1^2(z)/2\varkappa_1^2-{E_1^2(z)}/{2\varkappa_1^2}+{\alpha_2^2V_2^2(z)}/{2\varkappa_2^2}\,
-{E_2^2(z)}/{2\varkappa_2^2}$, its value far from the interface
is zero and therefore also at the interface. From the
continuity of the potential and of its derivative due to
eq.~(\ref{hered}) and (\ref{hered1}), we have that this is also
true at the wall just inside the system ${z} = 0_{+}$ thus
\begin{align}
\label{invariant}
\frac{\rho(0_{+})}{\rho_b}+\frac{\alpha_1^2}{2\varkappa_1^2}\,
\left[V_1(0_{+})\right]^2-\frac{1}{2\varkappa_1^2}E_1^2(0_{+})+&\frac{\alpha_2^2}{2\varkappa_2^2}\,
\left[V_2(0_{+})\right]^2-\frac{1}{2\varkappa_2^2}E_2^2(0_{+})=\frac{\rho(0_{+})}{\rho_b}.
\end{align}

As this quantity is constant we obtain the so-called contact
theorem
\begin{eqnarray}
\label{CT}
    {\beta P}={\rho}(0_{+}).
\end{eqnarray}
Thus, similar to the one-Yukawa case \cite{molphys}, in the MFA
we obtain the contact theorem as the consequence of the
existence of an invariant of the differential equations
corresponding to the bulk pressure.

\subsection{Density profiles}\label{mfa_density}
From (\ref{grad})-(\ref{hered1}) we obtain a set of five
differential equations with five unknown functions $\rho(z)$,
$E_1(z)$, $E_2(z)$, $V_1(z)$, $V_2(z)$:
\begin{align}
\label{logd}
\frac{\partial\rho(z)}{\partial z}&=\rho(z)\left[E_1(z)+E_2(z)\right],\\
\frac{\partial V_1(z)}{\partial z}&=-E_1(z),\\
\frac{\partial V_2(z)}{\partial z}&=-E_2(z),\\
\frac{\partial E_1(z)}{\partial z}&=-\alpha_1^2V_1(z)+\frac{\varkappa_1^2}{\rho_b}\rho(z),\\
\frac{\partial E_2(z)}{\partial z}&=-\alpha_2^2V_2(z)+\frac{\varkappa_2^2}{\rho_b}\rho(z)\label{logd3}.
\end{align}
These relations are first-order differential equations that
can be solved numerically.

From ($\ref{MFd}$) we have
\begin{align}
\label{f3}
\rho(z)=\rho_b\exp\big(-[V_1(z)-V_{1b}]-[V_2(z)-V_{2b}]\big),
\end{align}
where $V_i(z)$ and $V_{ib}$ are given by (\ref{f1}) and (\ref{f2}).

Similar to \cite{molphys} we can solve equation (\ref{f3}) in the linear approximation with the boundary condition set by the contact theorem.
This linear solution was obtained in \cite{preprint2013}.
Here we present only the final result
\begin{align}
\label{linearized_d}
\frac{\rho^{L}(z)}{\rho_b}&=1-\frac{1}{2}\frac{\left(\lambda_1^2-\alpha_2^2\right)}{\left(\lambda_1^2-\lambda_2^2\right)}
\left(-\frac{\varkappa_1^2}{\alpha_1^2}
+\frac{\lambda_2^2-\alpha_2^2-\varkappa_2^2}{\alpha_2^2}\right)\,e^{\displaystyle-\lambda_1 z}\\
&-\frac{1}{2}\frac{\left(\lambda_2^2-\alpha_2^2\right)}{\left(\lambda_1^2-\lambda_2^2\right)}\left(\frac{\varkappa_1^2}{\alpha_1^2}
-\frac{\lambda_1^2-\alpha_2^2-\varkappa_2^2}{\alpha_2^2}\right)\,e^{\displaystyle-\lambda_2 z},\nonumber
\end{align}
where
\begin{align}
\label{eigen}
\lambda_{1,2}^2=\frac{1}{2}\left(\varkappa_1^2+\alpha_1^2+\varkappa_2^2+\alpha_2^2\pm\sqrt{\left(\varkappa_1^2+
\alpha_1^2-\varkappa_2^2-\alpha_2^2\right)^2+4\varkappa_1^2\varkappa_2^2}\right).
\end{align}

\section{Fluctuation and correlation effects on density
profiles at the wall}

In the previous section we have considered mean field
equations, where the fluctuations are neglected. Here we take
them into account and therefore we have to expand the
Hamiltonian with respect to the mean field density
$\rho^{MFA}(\mathbf{r})$. For this aim we put
$\rho(\mathbf{r})=\rho^{MFA}(\mathbf{r})+\delta\rho(\mathbf{r})$.

\subsection{Expansion of the Hamiltonian}
Expansion of the Hamiltonian around the mean field density
$\rho^{MFA}(\mathbf{r})$ gives
\begin{align}
&\beta H[\rho]=\beta H\left[\rho^{MFA}\right] +
\int\delta\rho(\mathbf{r_1})\frac{\delta\beta H}
{\delta(\delta\rho(\mathbf{r_1}))}\bigg\vert_{\rho^{MFA}}d\mathbf{r_1}+
 \\\nonumber
 &\frac{1}{2}\int\delta\rho(\mathbf{r_1})\delta\rho(\mathbf{r_2})\frac{\delta^2\beta H}
 {\delta(\delta\rho(\mathbf{r_1}))\delta(\delta\rho(\mathbf{r_2})})
 \bigg\vert_{\rho^{MFA}}d\mathbf{r_1}
 d\mathbf{r_2}+\\
&\sum\limits_{n\geq 3}(-1)^n \frac{(n-1)!}
{n!}\int\delta\rho(\mathbf{r_1})\,...\,
\delta\rho(\mathbf{r_n})\frac{\delta^n\beta H}
 {\delta(\delta\rho(\mathbf{r_1}))\,...\,\delta(\delta\rho(\mathbf{r_n})})\bigg\vert_{\rho^{MFA}}
  d\mathbf{r_1}...d\mathbf{r_n}.\nonumber
\end{align}
The first term is the Hamiltonian functional (\ref{eq1}) for
the mean field density
\begin{align}
\beta H&[\rho^{MFA}]=\int \rho^{MFA}(\mathbf{r_1})
\left(\ln\left[\rho^{MFA}(\mathbf{r_1})\Lambda^3\right]
-1\right)d\mathbf{r_1}\\\nonumber
&+\frac{\beta}{2}\int\nu(r_{12})\bigg[\rho^{MFA}(\mathbf{r_1})
\rho^{MFA}(\mathbf{r_2})-\rho^{MFA}(\mathbf{r}_1)\delta(\mathbf{r}_1-\mathbf{r}_2)\bigg]
d\mathbf{r_1}d\mathbf{r_2}.
    \end{align}
The linear term disappears as in the canonical formalism
fluctuations preserve the number of particles and
$\int\delta\rho(\mathbf{r})d\mathbf{r}=0$.

The quadratic term is
\begin{align}
\beta H_2[\rho]=\frac{1}{2}\int\delta\rho(\mathbf{r_1})\delta\rho(\mathbf{r_2})\left[\frac{\delta(\mathbf{r}_1-
\mathbf{r}_2)}{\rho^{MFA}(\mathbf{r_1})}+\beta\nu(r_{12})\right]d\mathbf{r_1}d\mathbf{r_2},
\end{align}
where the first term comes from the expansion of the
logarithmic term in the Hamiltonian.

Due to translational invariance parallel to the wall, we expand
the fluctuations of the density as
\begin{align}
\delta\rho(\mathbf{r})=\sum\limits_{\mathbf{K}}\delta\rho_{\mathbf{K}}(z)\,
e^{\displaystyle \text{i}\mathbf{K}\mathbf{R}},
\end{align}
where ${\mathbf{R}}$ is the vector component of ${\mathbf{r}}$
parallel to the wall, ${\mathbf{K}}$ is the wave vector in the
direction parallel to the wall.

The entropic term equals
\begin{align}
\beta H_2^{entr}\left[\rho_{\mathbf{K}}(z)\right]&=\frac{1}{2}\int
\frac{\delta\rho^2(\mathbf{r})}{\rho^{MFA}(z)}\,d\mathbf{r}\\\nonumber
&=\frac{1}{2}
\sum\limits_{\mathbf{K},\mathbf{K^{'}}}\int\frac{\delta\rho_{\mathbf{K}}(z)
\delta\rho_{\mathbf{K}^{'}}(z)}{\rho^{MFA}(z)}\,e^{\displaystyle\text{i}\mathbf{R}(\mathbf{K}+\mathbf{K}^{'})}
d\mathbf{R}dz
\\
&=\frac{S}{2}\sum\limits_{\mathbf{K}}
\int dz_1dz_2\,\delta\rho_{\mathbf{K}}(z_1)
\delta\rho_{-\mathbf{K}}(z_2)\,\frac{\delta(z_1-z_2)}{\rho^{MFA}(z)},\nonumber
\end{align}
where $S$ is the surface area.

The interaction term gives
\begin{align} \beta H_2^{int}\left[\rho_{\mathbf{K}}(z)\right]=
\frac{S\,{\beta}}{2}\sum\limits_{\mathbf{K}}\int d z_1\int d z_2
\delta\rho_{\mathbf{K}}(z_1)
\delta\rho_{-\mathbf{K}}(z_2)\,
\nu\left(\mathbf{K},\vert z_1-z_2\vert\right),
\end{align}
where $\nu\left(\mathbf{K},\vert z_1-z_2\vert\right)=\int
d\mathbf{R}_{12}\,\nu(r_{12})\,\exp{\left(-i\mathbf{K}\mathbf{R}_{12}\right)}$.

Finally, for the quadratic term of the Hamiltonian we obtain
\begin{align}
\label{quadratic}
&\beta H_2[\rho]=\\
\nonumber &\frac{S}{2}\sum\limits_{\mathbf{K}}\int d z_1\int d z_2
\,\delta\rho_{\mathbf{K}}(z_1)
\delta\rho_{-\mathbf{K}}(z_2)\,\left[\frac{\delta\left(z_1-z_2\right)}{\rho^{MFA}(z_1)}
+\beta\nu\left(\mathbf{K},\vert z_1-z_2\vert\right)\right].
\end{align}

\subsection{Thermodynamic properties: free energy, pressure, and
chemical potential}
We start our calculations from
consideration of thermodynamic properties of the fluid in the
bulk.

The free energy is
\begin{eqnarray}
 {\beta F} = -\ln\left[ \int \mathcal{D}\rho\; e^{-\beta H[\rho]}\right].
\end{eqnarray}
In order to calculate the functional integral using the
Gaussian integrals with such a Hamiltonian, it is necessary to
have the quadratic term in a diagonal form. For bulk properties
such as the Helmholtz free energy we can expand density on the
Fourier components
\begin{align}
\delta\rho(\mathbf{r})=\sum\limits_{\mathbf{k}}\delta\rho_{\mathbf{k}}\,
e^{\displaystyle \text{i}\mathbf{k}\mathbf{r}}.
\end{align}
In this basis the quadratic Hamiltonian is
\begin{align}
\beta H_2[\rho]=\frac{V}{2\rho_b}\sum\limits_{\mathbf{k}>0}
\delta\rho_{\mathbf{k}}
\delta\rho_{-\mathbf{k}}
\left[1+\frac{\varkappa_1^2}{\mathbf{k}^{2}+\alpha_1^2}
+\frac{\varkappa_2^2}{\mathbf{k}^{2}+\alpha_2^2}\right]
\end{align}
and after integration the excess free energy equals
\begin{align}
\label{eq2.52+}
&\beta F^{ex}=\beta (F-F^{id})=\\
\nonumber&\rho_b V\frac{\varkappa_1^2}{2\alpha_1^2}
+\rho_b V\frac{\varkappa_1^2}{2\alpha_1^2}
+\frac{1}{2}\,\sum\limits_{\rm
\mathbf{k}}^{}\,\ln\left[1+\rho_b\,{\nu}(k)\right]-\frac{1}{2}\,\rho_b\,\sum\limits_{\rm
\mathbf{k}}^{}\,{\nu}(k),
\end{align}
where
\begin{equation}
{\nu}(k)=\frac{4\pi\beta A_1}{k^2+\alpha_1^2}+\frac{4\pi\beta A_2}{k^2+\alpha_2^2}
\end{equation}
is the Fourier transform of the interaction potential
(\ref{pot1}) multiplied by $\beta$.

The first and the second terms on the right-hand side of
(\ref{eq2.52+}) are mean field contributions with the other two
terms coming from Gaussian fluctuations. In order to calculate
the third and the fourth terms we switch from summation to
integration and then integrate by parts
\begin{align}
\label{eq2.53i}
\beta F^{fluct}&=\frac{1}{2}\,\sum\limits_{\rm
\mathbf{k}}^{}\,\ln\left[1+\rho_b{\nu}(k)\right]\,-\,\frac{1}{2}\,\rho_b\,\sum\limits_{\rm
\mathbf{k}}^{}\,{\nu}(k)\,\\\nonumber
&=\frac{{\rho_b^2}{V}}{12\pi^2}\,\int\limits_{0}^{\infty}k^3dk
\frac{{\nu}(k)}{1+\rho_b{\nu}(k)}\frac{d\,{\nu(k)}}{dk}.
\end{align}
For further calculations it is helpful to express parameters
$\varkappa_1^2$ and $\varkappa_2^2$ in terms of $\lambda_1$ and
$\lambda_2$. From (\ref{eigen}) we have
\begin{align}
\label{A_1}
\varkappa_1^2=\frac{\left(\alpha_1^2-\lambda_1^2\right)\left(\alpha_1^2-\lambda_2^2\right)}
{\alpha_2^2-\alpha_1^2},\qquad
\varkappa_2^2=\frac{\left(\alpha_2^2-\lambda_1^2\right)\left(\alpha_2^2-\lambda_2^2\right)}
{\alpha_1^2-\alpha_2^2}.
\end{align}
Using identities (\ref{A_1}), after integration we obtain
\begin{align}
\label{F}
\frac{\beta F^{ex}}{V}=\,&\frac{\rho_b}{2}\left(\frac{\varkappa_1^2}{\alpha_1^2}
+\frac{\varkappa_2^2}{\alpha_2^2}\right)-{\frac {1}{12\pi }}({\lambda_{{1}}}^{3}+{\lambda_{{2}}}^{3})-
{\frac {1}{24\pi }}({\alpha_{{1}}}^{3}+{\alpha_{{2}}}^{3})\\
&+{\frac {1}{8\pi }}\left( {\lambda_{{1}}}^{2}+{\lambda_{{2}}}^{2}
\right)  \left( \alpha_{{1}}+\alpha_{{2}} \right)- \,\frac
{1}{8\pi}\frac{\left( { \lambda_{{1}}}^{2}+\alpha_{{1}}\alpha_{{2}}
\right)  \left( {\lambda_{{2 }}}^{2}+\alpha_{{1}}\alpha_{{2}}
\right)}{\alpha_{{1}}+\alpha_{{2}}}.\nonumber
\end{align}
The pressure can be found from the free energy as
\begin{align}
\beta P =-\beta\frac{\partial F}{\partial V}\bigg\vert_{T,N}.
\end{align}
Differentiation of (\ref{F}) with respect to volume gives the
fluctuation part of the bulk pressure as
\begin{eqnarray}
\label{pressure2}
\beta P^{fluct}=
=\,\frac{\rho_b^2}{12\pi^2}\,\int\limits_{0}^{\infty}k^3dk
\frac{{\nu}(k)}{\left[1+\rho_b{\nu}(k)\right]^2}
\frac{d\,{\nu}(k)}{dk}.\quad
\end{eqnarray}
After integration and due to identities (\ref{A_1}) the excess
pressure equals
\begin{align}
\label{P}
&\beta P^{ex}=\frac{\rho_b}{2}\,\left(\frac{\varkappa_1^2}{\alpha_1^2}
+\frac{\varkappa_2^2}{\alpha_2^2}\right)-{\frac {1}{24\pi }}
({\lambda_{{1}}}^{3}+{\lambda_{{2}}}^{3})-\frac {1}{12\pi }({\alpha_{{1}}}^{3}+{\alpha_{{2}}}^{3})\\
&+
\frac {1}{8\pi } \left( {\alpha_{{1}}}^{2}+{\alpha_{{2}}}^{2} \right)
\left(\lambda_{{1}}+\lambda_{{2}} \right)- \frac {1}{8\pi} \frac{1}{\lambda_{{1}}+
\lambda_{{2}}} \left({\alpha_{{1}}}^{2}+\lambda_{{1}}\lambda_{{2}} \right)
\left( {\alpha_{{2}}}^{2}+\lambda_{{1}}\lambda_{{2}} \right).\nonumber
\end{align}
Finally, the excess chemical potential can be derived directly
from (\ref{F}) and (\ref{P}) as
$\mu^{ex}=\left(F^{ex}+P^{ex}V\right)/N$ giving
\begin{align}
\label{mu}
\beta\mu^{ex}=\,&\frac{\varkappa_1^2}{\alpha_1^2}
+\frac{\varkappa_2^2}{\alpha_2^2}
-\frac{1}{8\pi\rho_b}(\lambda_1^3+\lambda_2^3)-\frac{1}{8\pi\rho_b}(\alpha_1^3+\alpha_2^3)\\
&+\frac{1}{8\pi\rho_b}(\lambda_1^2+\lambda_2^2)(\alpha_1+\alpha_2)+\frac{1}{8\pi\rho_b}(\alpha_1^2+\alpha_2^2)
(\lambda_1+\lambda_2)\nonumber\\\nonumber
&-\frac{1}{8\pi\rho_b}\frac{(\lambda_1^2+\alpha_1\alpha_2)(\lambda_2^2+\alpha_1\alpha_2)}{\alpha_1+\alpha_2}-
\frac{1}{8\pi\rho_b}\frac{(\alpha_1^2+\lambda_1\lambda_2)(\alpha_2^2+\lambda_1\lambda_2)}{\lambda_1+\lambda_2}.
\end{align}

\subsection{Correlation function}
The expression for the pair correlation function
$h(\mathbf{r}_1,\mathbf{r}_2)$ is \cite{hansen}
\begin{align}
h(\mathbf{r}_1,\mathbf{r}_2)\langle\rho(\mathbf{r}_1)\rangle\langle\rho(\mathbf{r}_2)\rangle=
\langle\delta\rho(\mathbf{r}_1)\delta\rho(\mathbf{r}_2)\rangle-\delta\left(\mathbf{r}_1-\mathbf{r}_2\right)
\langle\rho(\mathbf{r}_1)\rangle.
\end{align}
In $\mathbf{K}$-space this expression reads
\begin{align}
\label{dens-dens}
\frac{1}{S}\bigg[\rho^{MFA}(z_1)\rho^{MFA}(z_2)\,h({K},z_1z_2)+
&\rho^{MFA}(z_1)\delta(z_1-z_2)\bigg]\\\nonumber
&=\langle\delta\rho_{\mathbf{K}}(z_1)\delta\rho_{-\mathbf{K}}(z_2)\rangle,
\end{align}
where
\begin{align}
h({K},z_1z_2)=\int d\mathbf{R}_{12}\,h(R_{12},z_1,z_2)\,
\exp\left(i\mathbf{K}\mathbf{R}_{12}\right).
\end{align}
The right-hand side of equation (\ref{dens-dens}) can be
calculated from expression (\ref{quadratic}) and gives the
inverse Hamiltonian matrix $\beta
H_2^{-1}\left[\rho_{\mathbf{K}}(z)\right]/2$
\begin{align}
\nonumber\langle\delta\rho_{\mathbf{K}}(z_1)\delta\rho_{-\mathbf{K}}(z_2)\rangle&=
\frac{\int D(\delta\rho_{\mathbf{K}}(z))\delta\rho_{\mathbf{K}}(z_1)\delta\rho_{-\mathbf{K}}(z_2)
\exp\left(-\beta H_2[\rho_{\mathbf{K}}(z)]\right)}{\int D(\delta\rho_{\mathbf{K}}(z))
\exp\left(-\beta H_2[\rho_{\mathbf{K}}(z)]\right)}\\&=\frac{1}{2}\beta
H_2^{-1}\left[\rho_{\mathbf{K}}(z)\right].
\label{defin}
\end{align}
Hence the product of the Hamiltonian matrix and the matrix on
the left-hand side of (\ref{dens-dens}) yields unity, so we
have
\begin{align}
\int dz_3&\bigg(\bigg[\rho^{MFA}(z_1)\rho^{MFA}(z_3)\,h({K},z_1z_3)+
\rho^{MFA}(z_1)\delta(z_1-z_3)\bigg]\nonumber\\
&\;\;\;\;\;\;\;\;\;\;\;\;\left[\frac{\delta\left(z_3-z_2\right)}{\rho^{MFA}(z_3)}
+\beta\nu\left(\mathbf{K},\vert z_3-z_2\vert\right)\right]\bigg)=\delta(z_1-z_2),
\end{align}
or
\begin{align}
\label{conv}
h({K},z_1,z_2)+\int dz_3\rho^{MFA}(z_3)h({K},z_1,z_3)
\beta&\nu\left(\mathbf{K},\vert z_3-z_2\vert\right)\\\nonumber&=-\beta\nu\left(\mathbf{K},\vert z_1-z_2\vert\right).
\end{align}
Relation (\ref{conv}) is a convolution-type equation. It can be
reduced to the so-called Riemann problem \cite{hakhov} if we
assume the density profile to be a step-function. In this
approximation $\rho^{MFA}(z)=0$ for $z<0$ and
$\rho^{MFA}(z)=\rho_b$ for $z>0$.

Due to the spatial non-homogeneousness of the system we
introduce one-sided pair correlation functions
$h_{\pm}({R_{12}},z_1,z_2)$ such that
\begin{eqnarray}
h(R_{12},z_1,z_2)&=&h_{+}(R_{12},z_1,z_2)-h_{-}(R_{12},z_1,z_2),\nonumber\\
h_{+}(R_{12},z_1,z_2)&=&\left\{\begin{array}{ll}
h(R_{12},z_1,z_2), & z_1>0, \\
0, & z_1<0,
\end{array} \right. \\
h_{-}(R_{12},z_1,z_2)&=&\left\{\begin{array}{ll}
0, & z_1>0, \\
-h(R_{12},z_1,z_2),& z_1<0.
\end{array} \right. \nonumber\\
\label{eq2.28}\nonumber.
\end{eqnarray}
The function $h({K},z_1,z_2)$ can then be presented as the
difference of one-sided functions $h_{\pm}({K},z_1,z_2)$ such
that
\begin{eqnarray}
h({K},z_1,z_2)&=&h_{+}({K},z_1,z_2)-h_{-}({K},z_1,z_2),\nonumber\\
h_{+}({K},z_1,z_2)&=&\left\{\begin{array}{ll}
h({K},z_1,z_2), & z_1>0, \\
0, & z_1<0,
\end{array} \right. \\
h_{-}({K},z_1,z_2)&=&\left\{\begin{array}{ll}
0, & z_1>0, \\
-h({K},z_1,z_2),& z_1<0.
\end{array} \right. \nonumber\\
\label{Kzz}\nonumber
\end{eqnarray}
Equation (\ref{conv}) now reads
\begin{align}
\label{conv1}
\nonumber h_{+}({K},z_1,z_2)-h_{-}({K},z_1,z_2)+\rho_b\int\limits_0^{\infty}dz_3 &h_{+}({K},z_1,z_3)
\beta\nu\left(\mathbf{K},\vert z_3-z_2\vert\right)\\
&=-\beta\nu\left(\mathbf{K},\vert z_1-z_2\vert\right).
\end{align}
Expanding the functions $h_{\pm}({K},z_1,z_2)$ and
$\nu\left(\mathbf{K},\vert z_1-z_2\vert\right)$ on Fourier
harmonics with respect to the wave vector $\mu$ in the
direction perpendicular to the wall and switching from
summation to integration we obtain
\begin{align}
\label{eq}
\left[1+\frac{\varkappa_1^2}{{K}^{2}+\mu_1^{2}+\alpha_1^2}
+\frac{\varkappa_2^2}{{K}^{2}+\mu_1^{2}+\alpha_2^2}\right]\,h_{+}({K},\mu_1\mu_2)-h_{-}({K},\mu_1\mu_2)\\
=
-\left(\frac{4\pi\beta A_1}{K^2+\mu_2^2+\alpha_1^2}+\frac{4\pi\beta A_2}{K^2+\mu_2^2+\alpha_2^2}\right)
\delta(\mu_1+\mu_2),\nonumber
\end{align}
where
\begin{align}
{h}_{\pm}(K,\mu_1,\mu_2)\,=\,
\int\limits_{\cal{S}}\mathbf{dR_{12}}\mbox{\large{e}}^{i\mathbf{KR_{12}}}
\int\limits_{-\infty}^{\infty}dz_1\mbox{\large{e}}^{\displaystyle i\mu_1z_1}
\int\limits_{-\infty}^{\infty}dz_2\mbox{\large{e}}^{\displaystyle i\mu_2z_2}
h_{\pm}(R_{12},z_1z_2)
\end{align}
and we have used the relation
\begin{align}
\int\limits_{-\infty}^{\infty}dz_1\mbox{\large{e}}^{\displaystyle i\mu_1z_1}
\int\limits_{-\infty}^{\infty}&dz_2\mbox{\large{e}}^{\displaystyle i\mu_2z_2}
\beta\nu\left(\mathbf{K},\vert z_1-z_2\vert\right)=\\\nonumber
&=-\left(\frac{4\pi\beta A_1}{K^2+\mu_2^2+\alpha_1^2}+\frac{4\pi\beta A_2}{K^2+\mu_2^2+\alpha_2^2}\right)
\delta(\mu_1+\mu_2).
\end{align}
Equation (\ref{eq}) is known as the Riemann problem
\cite{hakhov}. Using the technique proposed in
\cite{soviak,preprint2013} we solve this problem for
$h_{+}({K},\mu_1\mu_2)$ (refer to Appendix \ref{app:eigen} for
the details of calculation) and obtain
\begin{eqnarray}
&& h_{+}(K,\mu_1,\mu_2)=\nonumber\\
&&-\frac{1}{\rho_b}\frac{{\varkappa_1^2}{(\mu_2^2+\alpha_2^2(K))}
+{\varkappa_2^2}{(\mu_2^2+\alpha_1^2(K))}}{(\mu_2-i\alpha_{1}(K))(\mu_2-i\alpha_{2}(K))
(\mu_2+i\lambda_{1}(K))(\mu_2+i\lambda_{2}(K))}\nonumber\\
&&\qquad\qquad\qquad\frac{(\mu_1+i\alpha_{1}(K))(\mu_1+i\alpha_{2}(K))}
{(\mu_1+i\lambda_{1}(K))(\mu_1+i\lambda_{2}(K))}\,\delta_{+}(\mu_1+\mu_2),
\label{eq2.40!}
\end{eqnarray}
where $\delta_{+}(\mu_1+\mu_2)$ is a one-sided delta-function.

The expression for the correlation function in
$\mathbf{r}$-space can be presented as the sum of the homogeneous bulk part $h_+^b(r_{12})$ and the inhomogeneous surface part
 $h_{+}^{inh}(r_{12},z_1z_2)$
 \begin{align}
h_{+}(r_{12},z_1,z_2)=h_+^b(r_{12})+h_+^{inh}(r_{12},z_1z_2),
\end{align}
where
\begin{align}
\label{h-bulk}
h_+^b(r_{12})=&\beta
\frac{{A_1}{(\lambda_2^2-\alpha_2^2)}
+A_2{(\lambda_2^2-\alpha_1^2)}}{(\lambda_1^2-\lambda_2^2)}
\frac{\mbox{\large{e}}^{-\lambda_2r_{12}}}{r_{12}}\,-\\
&\beta\frac{{A_1}{(\lambda_1^2-\alpha_2^2)}
+A_2{(\lambda_1^2-\alpha_1^2)}}{(\lambda_1^2-\lambda_2^2)}
\frac{\mbox{\large{e}}^{-\lambda_1r_{12}}}{r_{12}},\nonumber
\end{align}
\begin{align}
\label{h_r}
&h_{+}^{inh}(r_{12},z_1,z_2)=\\
&-\beta
\int\limits_{0}^{\infty}2K\,J_{0}(KR_{12})\,dK \left\{\frac{{A_1}{(\lambda_2^2-\alpha_2^2)}
+A_2{(\lambda_2^2-\alpha_1^2)}}{2\lambda_2(K)(\lambda_1(K)-\lambda_2(K))^2}\right.\nonumber\\
&\qquad\qquad\qquad\left.
\frac{(\lambda_2(K)-\alpha_1(K))(\lambda_2(K)-\alpha_2(K))}
{(\lambda_2(K)+\alpha_1(K))(\lambda_2(K)+\alpha_2(K))}
\mbox{\large{e}}^{-\lambda_2(K)(z_1+z_2)}\right.\nonumber\\
&-\left.\frac{{A_1}{(\lambda_1^2-\alpha_2^2)}
+A_2{(\lambda_1^2-\alpha_1^2)}}
{(\lambda_1(K)+\lambda_2(K))(\lambda_1(K)-\lambda_2(K))^2}\right.\nonumber\\
&\left.\qquad\qquad\qquad\frac{(\lambda_2(K)-\alpha_1(K))(\lambda_2(K)-\alpha_2(K))}
{(\lambda_1(K)+\alpha_1(K))(\lambda_1(K)+\alpha_2(K))}
\mbox{\large{e}}^{-\lambda_2(K)z_1-\lambda_1(K)z_2}\right.\nonumber\\
&-\left.\frac{{A_1}{(\lambda_2^2-\alpha_2^2)}
+A_2{(\lambda_2^2-\alpha_1^2)}}
{(\lambda_1(K)+\lambda_2(K))(\lambda_1(K)-\lambda_2(K))^2}\right.\nonumber\\
&\left.\qquad\qquad\qquad\frac{(\lambda_1(K)-\alpha_1(K))(\lambda_1(K)-\alpha_2(K))}
{(\lambda_2(K)+\alpha_1(K))(\lambda_2(K)+\alpha_2(K))}
\mbox{\large{e}}^{-\lambda_1(K)z_1-\lambda_2(K)z_2}\right.\nonumber\\
&+\left.\frac{{A_1}{(\lambda_1^2-\alpha_2^2)}
+A_2{(\lambda_1^2-\alpha_1^2)}}{2\lambda_1(K)(\lambda_1(K)-\lambda_2(K))^2}\right.\nonumber\\
&\left.\qquad\qquad\qquad\frac{(\lambda_1(K)-\alpha_1(K))(\lambda_1(K)-\alpha_2(K))}
{(\lambda_1(K)+\alpha_1(K))(\lambda_1(K)+\alpha_2(K))}
\mbox{\large{e}}^{-\lambda_1(K)(z_1+z_2)}\right\},\nonumber
\label{eq2.46}
\end{align}
\begin{align}
\alpha_i(K)=\sqrt{\alpha_i^2+K^2};\quad\quad \lambda_i(K)=\sqrt{\lambda_i^2+K^2},
\end{align}
and $J_{0}(KR_{12})$ is the Bessel function of the first kind.

As we see from (\ref{h-bulk}), $\lambda_1$ and
$\lambda_2$ play the role of
 parameters characterizing the screening of the repulsive
 and the attractive interactions
 respectively.

%

\subsection{Density profile}
In the Gaussian approximation the inhomogeneous density profile
can be written as the sum of the mean field profile
$\rho^{MFA}(z)$ and the quadratic fluctuation term
$\rho^{fluct}(z)$
\begin{align}
\label{twoparts}
{\rho(z)}={\rho^{MFA}(z)}+{\rho^{fluct}(z)}.
\end{align}
The contribution of quadratic fluctuations to the profile
corresponds to the one-particle irreducible diagram in the
field theory \cite{Amit,Zinn-Justin} and can be found as:
\begin{align}
\frac{\rho^{fluct}(z_1)}{\rho_b}=\frac{1}{2}\left[h_{+}(R,z_1,z_2)-h^b_{+}(R,z_1,z_2)\right]
\mathop{\bigg\vert_{z_2\rightarrow z_1}}\limits_{R\rightarrow 0},
\end{align}
where calculating the inhomogeneous profile we have subtracted
the homogeneous bulk part.

As a result
\begin{align}
\label{density_profile}
&\frac{\rho^{fluct}(z_1)}{\rho_b}=-\frac{1}{8\pi\rho_b}\int\limits_{0}^{\infty}K\,dK
\bigg\{\frac{{\varkappa_1^2}{(\lambda_{1}^2-\alpha_2^2)}
+\varkappa_2^2{(\lambda_{1}^2-\alpha_1^2)}}{\lambda_{1}(K)(\lambda_{2}(K)-\lambda_{1}(K))^2}\times\\\nonumber
&\;\;\;\;\;\;\;\;\;\;\;\;\;\frac{(\lambda_{1}(K)-\alpha_1(K))(\lambda_{1}(K)-\alpha_2(K))}
{(\lambda_{1}(K)+\alpha_1(K))(\lambda_{1}(K)+\alpha_2(K))}
\,\mbox{\large{e}}^{-2\lambda_{1}(K)z_1}\nonumber\\\nonumber\\
&-2\bigg[\frac{{\varkappa_1^2}{(\lambda_{2}^2-\alpha_2^2)}
+\varkappa_2^2{(\lambda_{2}^2-\alpha_1^2)}}
{(\lambda_{2}(K)+\lambda_{1}(K))(\lambda_{2}(K)-\lambda_{1}(K))^2}\times\nonumber\\
&\;\;\;\;\;\;\;\;\;\;\;\;\;\frac{(\lambda_{1}(K)-\alpha_1(K))(\lambda_{1}(K)-\alpha_2(K))}
{(\lambda_{2}(K)+\alpha_1(K))(\lambda_{2}(K)+\alpha_2(K))}\nonumber\\\nonumber
&+\frac{{\varkappa_1^2}{(\lambda_{1}^2-\alpha_2^2)}
+\varkappa_2^2{(\lambda_{1}^2-\alpha_1^2)}}
{(\lambda_{2}(K)+\lambda_{1}(K))(\lambda_{2}(K)-\lambda_{1}(K))^2}\times\\
&\;\;\;\;\;\;\;\;\;\;\;\;\;\frac{(\lambda_{2}(K)-\alpha_1(K))(\lambda_{2}(K)-\alpha_2(K))}
{(\lambda_{1}(K)+\alpha_1(K))(\lambda_{1}(K)+\alpha_2(K))}\bigg]\,
\mbox{\large{e}}^{-[\lambda_{1}(K)+\lambda_{2}(K)]z_1}\nonumber\\\nonumber\\
&+\frac{{\varkappa_1^2}{(\lambda_{2}^2-\alpha_2^2)}
+\varkappa_2^2{(\lambda_{2}^2-\alpha_1^2)}}{\lambda_{2}(K)(\lambda_{2}(K)-\lambda_{1}(K))^2}\times\nonumber\\
&\;\;\;\;\;\;\;\;\;\;\;\;\;\frac{(\lambda_{2}(K)-\alpha_1(K))(\lambda_{2}(K)-\alpha_2(K))}
{(\lambda_{2}(K)+\alpha_1(K))(\lambda_{2}(K)+\alpha_2(K))}
\,\mbox{\large{e}}^{-2\lambda_{2}(K)z_1}\bigg\}\nonumber.
\end{align}

\subsection{Contact theorem}

In Section \ref{contact_mf} we have shown the validity of the
contact theorem in the mean field approximation. Here we will
show that for the considered model the contact theorem is also
satisfied when the fluctuations are taken into account.

Setting $z_1=0$ in expression (\ref{density_profile}) and using
identities (\ref{A_1}), we obtain the contact value of density
\begin{align}
\label{contact2}
\rho^{fluct}(0_{+})=&\frac{1}{4\pi}\,\int\limits_0^{\infty}KdK\bigg[\alpha_1(K)+\alpha_2(K)-\frac{1}{2}\,
{\left[\lambda_1(K)+\lambda_2(K)\right]}\\\nonumber
&-\frac{1}{2}\,
\frac{\left[\alpha_1^2(K)+\lambda_1(K)\lambda_2(K)\right]\,\left[\alpha_2^2(K)+\lambda_1(K)\lambda_2(K)\right]}
{\lambda_1(K)\lambda_2(K)\left[\lambda_1(K)+\lambda_2(K)\right]}\,\bigg].
\end{align}
Going back to expression (\ref{pressure2}) for the pressure we
can calculate the fluctuation part of the pressure using the
cylindrical coordinate system instead of the spherical one.
Then we have
\begin{align}
\beta P&^{fluct}=\,\frac{\rho_b^2}{12\pi^2}\,\int\limits_{0}^{\infty}k^3dk
\frac{{\nu}(k)}{\left[1+\rho{\nu}(k)\right]^2}
\frac{d\,{\nu}(k)}{dk}\\\nonumber
&=-\frac{1}{2\pi^2}\,\int\limits_{0}^{\infty}KdK
\int\limits_{-\infty}^{\infty}\mu^2\,d\mu\,\frac{\bigg[\varkappa_1^2\left(\mu^2+\alpha_2^2(K)\right)+
\varkappa_2^2\left(\mu^2+\alpha_1^2(K)\right)\bigg]}{\left[\mu^2+\lambda_1^2(K)\right]^2
\left[\mu^2+\lambda_2^2(K)\right]^2}\,\\\nonumber
&\frac{\left[\varkappa_1^2\left(\mu^2+\alpha_2^2(K)\right)^2+
\varkappa_2^2\left(\mu^2+\alpha_1^2(K)\right)^2\right]}
{\left[\mu^2+\alpha_1^2(K)\right]\left[\mu^2+\alpha_2^2(K)\right]}.
\end{align}
After integration with respect to $\mu$ and taking into account
relations (\ref{A_1}) we obtain
\begin{align}
\beta P^{fluct}=\frac{1}{4\pi}\,\int\limits_0^{\infty}&KdK\bigg[\alpha_1(K)+\alpha_2(K)-\frac{1}{2}\,
{\left[\lambda_1(K)+\lambda_2(K)\right]}\\\nonumber
&-\frac{1}{2}\,
\frac{\left[\alpha_1^2(K)+\lambda_1(K)\lambda_2(K)\right]\,\left[\alpha_2^2(K)+\lambda_1(K)\lambda_2(K)\right]}
{\lambda_1(K)\lambda_2(K)\left[\lambda_1(K)+\lambda_2(K)\right]}\,\bigg],
\end{align}
which is exactly the expression (\ref{contact2}).

We have therefore proved the validity of the contact theorem
for the fluctuation term of the density profile.

\subsection{Adsorption}
We can also calculate the adsorption coefficient defined as
\begin{align}
\label{ads}
\Gamma\,=\,\int\limits_{0}^{\infty}dz\left[\rho(z)\,-\,\rho_b\right]=\Gamma_{MFA}+\Gamma_{fluct}
\end{align}
according to different approximations of the mean field density
profile presented in Section {\ref{mfa_density}.

Hence the exact mean field contribution can be determined only
numerically.

The linearized equation (\ref{linearized_d}) gives
\begin{align}
\label{gamma_lin}
\Gamma_{MFA}^L=&-\frac{\rho_b}{2\lambda_1}\frac{\left(\lambda_1^2-\alpha_2^2\right)}{\left(\lambda_1^2-\lambda_2^2\right)}
\left(-\frac{\varkappa_1^2}{\alpha_1^2}
+\frac{\lambda_2^2-\alpha_2^2-\varkappa_2^2}{\alpha_2^2}\right)\\\nonumber
&-\frac{\rho_b}{2\lambda_2}\frac{\left(\lambda_2^2-\alpha_2^2\right)}{\left(\lambda_1^2-\lambda_2^2\right)}\left(\frac{\varkappa_1^2}{\alpha_1^2}
-\frac{\lambda_1^2-\alpha_2^2-\varkappa_2^2}{\alpha_2^2}\right).
\end{align}

For the fluctuation part of the adsorption coefficient due to
identities (\ref{A_1}) we obtain an analytical result
\begin{align}
\nonumber\Gamma_{fluct}&=\frac{1}{32\pi}\, \left( \lambda_{{1}}+\lambda_{{2}} \right) ^{2}
+\frac{1}{32\pi}\,(\alpha_1^2+\alpha_2^2)
-\frac{1}{16\pi}\, \left( \lambda_{{1}}+\lambda_{{2}}
 \right)  \left( \alpha_{{1}}+\alpha_{{2}} \right)\\
&-\frac{1}{16\pi}\,{\frac {\left( \lambda_{{2}}\lambda_{{1}}+{\alpha_{{2}}}^{2} \right)  \left(
\lambda_{{2}}\lambda_{{1}}+{\alpha_{{1}}}^{2} \right) }{ \left( \lambda_{
{1}}+\lambda_{{2}} \right) ^{2}}}+\frac{1}{16\pi}\,{\frac { \left( \alpha_{{1}}+
\alpha_{{2}} \right)  \left( \lambda_{{2}}\lambda_{{1}}+\alpha_{{2}}
\alpha_{{1}} \right) }{\lambda_{{1}}+\lambda_{{2}}}}\nonumber\\
&+\frac{1}{16\pi}\left(\lambda_1^2+\lambda_2^2-\alpha_1^2-\alpha_2^2\right)\,
\ln\left[\frac{(\lambda_2+\alpha_1)(\lambda_2+\alpha_2)}{2\lambda_2(\lambda_1+\lambda_2)}\right]\nonumber\\
&+
\frac{1}{16\pi}\frac{(\lambda_1^2-\alpha_1^2)(\lambda_1^2-\alpha_2^2)}{\lambda_2^2-\lambda_1^2}
\,\ln\left[\frac{\lambda_1}{\lambda_2}\frac{(\lambda_2+\alpha_1)
(\lambda_2+\alpha_2)}{(\lambda_1+\alpha_1)(\lambda_1+\alpha_2)}\right].
\label{gamma-f}
\end{align}

\section{Monte-Carlo simulations}

{\color{black}
In order to test an accuracy of the field theoretical results established above the Monte Carlo~(MC)\cite{frenkel.book} simulations were carried out.
A system of fluid particles interacting with the two-Yukawa potential~(\ref{pot1}) was considered in a rectangular simulation box.
A cutoff distance of the potential was chosen $r_c=12.0$. A minimum size of the simulation box was set at least twice larger than the cutoff distance.
A usual periodic boundary conditions along x, y and z directions were applied for the bulk fluid. However, to study a fluid near the hard wall a simulation box was confined between two walls orthogonal to z-axis and in this case the periodic boundary conditions were applied only in xy-plane.
A distance between walls, $L_z$ , was taken large enough to form a wide layer of the bulk phase in the middle of the box~($L_z=37.0$).
A number of fluid particles depended on the considered densities ($\rho ^{*}=\rho_{b}/\alpha_{2}^{3}=0.1, 0.2, 0.3$) and it varied in the range of $N=6000-12000$.
At each simulation step $N$ trial movements of particles were performed. To speed up the simulations, the linked cell list
algorithm was employed\cite{frenkel.book}. The density profiles of a confined fluid, $\rho (z)$, were calculated and averaged over $100000$ simulation step,
while the pair distribution functions of a bulk fluid, $g(r)$, were averaged over $10000$ steps. The model was studied for the different ratios of parameters $A_1/\vert A_2\vert $ and $\alpha_1/\alpha_2$ in the temperature region of $T^{*}=T/(\alpha_2\vert A_2\vert)=0.5-1.5$.
}

\section{Results and discussion}

The properties of the considered two-Yukawa fluid are defined by four non-dimensional parameters:
$\rho ^{*}=\rho_{b}/\alpha_{2}^{3}$, $T^{*}=-1/(\beta A_{2}\alpha_{2} )$, $\omega=A_{1}/\vert A_{2}\vert$ and $\tau=\alpha_{1}/\alpha_{2} $. The first two parameters are non-dimensional density and temperature respectively. The last two parameters are connected with the form of interparticle interaction. Below we will consider three types of models with $\omega =3,\tau =1.498$; $\omega =2,\tau =1.35$ and $\omega =2.5,\tau =1.355$. The forms of interparticle interaction corresponding to these three cases are presented in Fig. \ref{fig1}.

\begin{figure}
\includegraphics[scale=0.3]{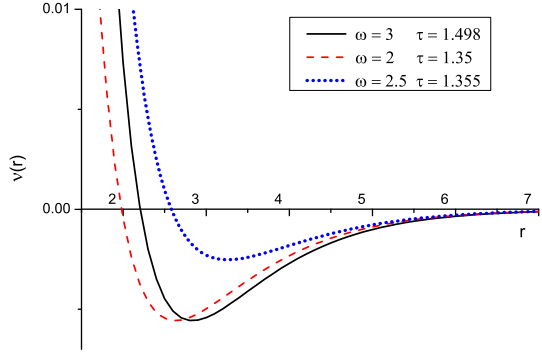}
\caption{Pair interaction potential (\ref{pot1}) for different values of $\omega$ and $\tau$.}
\label{fig1}
\end{figure}

We begin presentation of our results with the discussion of the bulk pair distribution function.

\subsection{Bulk pair distribution function}

The bulk pair distribution function (PDF) in the considered Gaussian approximation can be presented in the form
\begin{align}
\label{g}
g^{b}(r)=1+h_{+}^{b}(r),
\end{align}
where $h_{+}^{b}(r)$ is given by equation (\ref{h-bulk}). However, at small distances $h_{+}^{b}(r)$ is of Coulombic form and $h_{+}^{b}(r)\rightarrow -\infty$ when $r\rightarrow 0$. In order to avoid this non-physical behavior of $g^{b}(r)$ we can use the exponential form
\begin{align}
\label{g-exp}
g^{b}(r)=\exp\left[h_{+}^{b}(r)\right]
\end{align}
instead of the form (\ref{g}).

The behavior of the PDF for a model with $\omega =2,\tau =1.35, \rho^*=0.1$ at different temperatures is presented in Fig. \ref{fig2}. As we can see the exponential form (\ref{g-exp}) ensures the correct behavior of $g^b(r)$ at small distances and reproduces very well the results given by expression (\ref{g}) at large distances. These results are also in very good agreement with the computer simulations data.
\begin{figure}
{\includegraphics[scale=0.24]{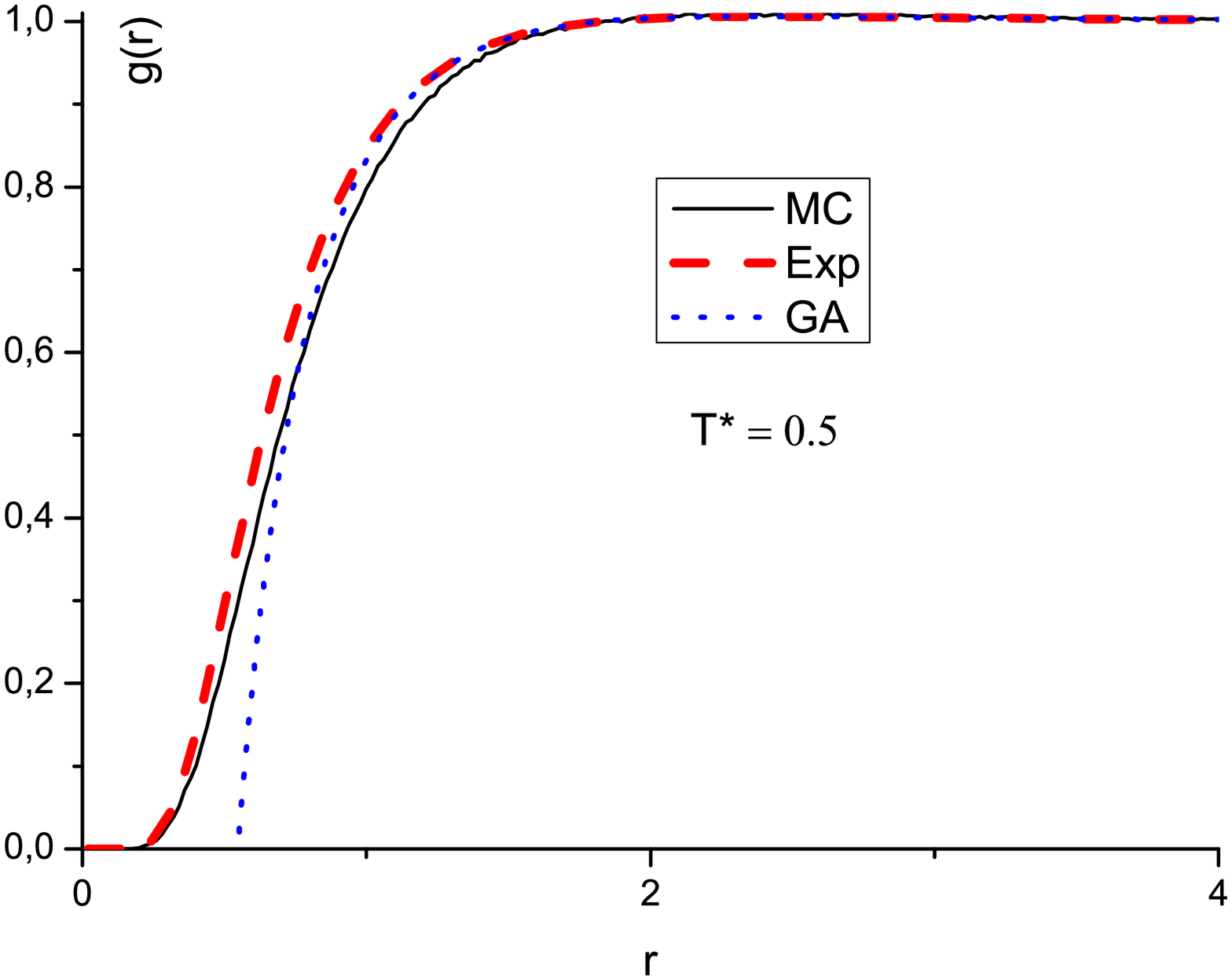}}
{\includegraphics[scale=0.24]{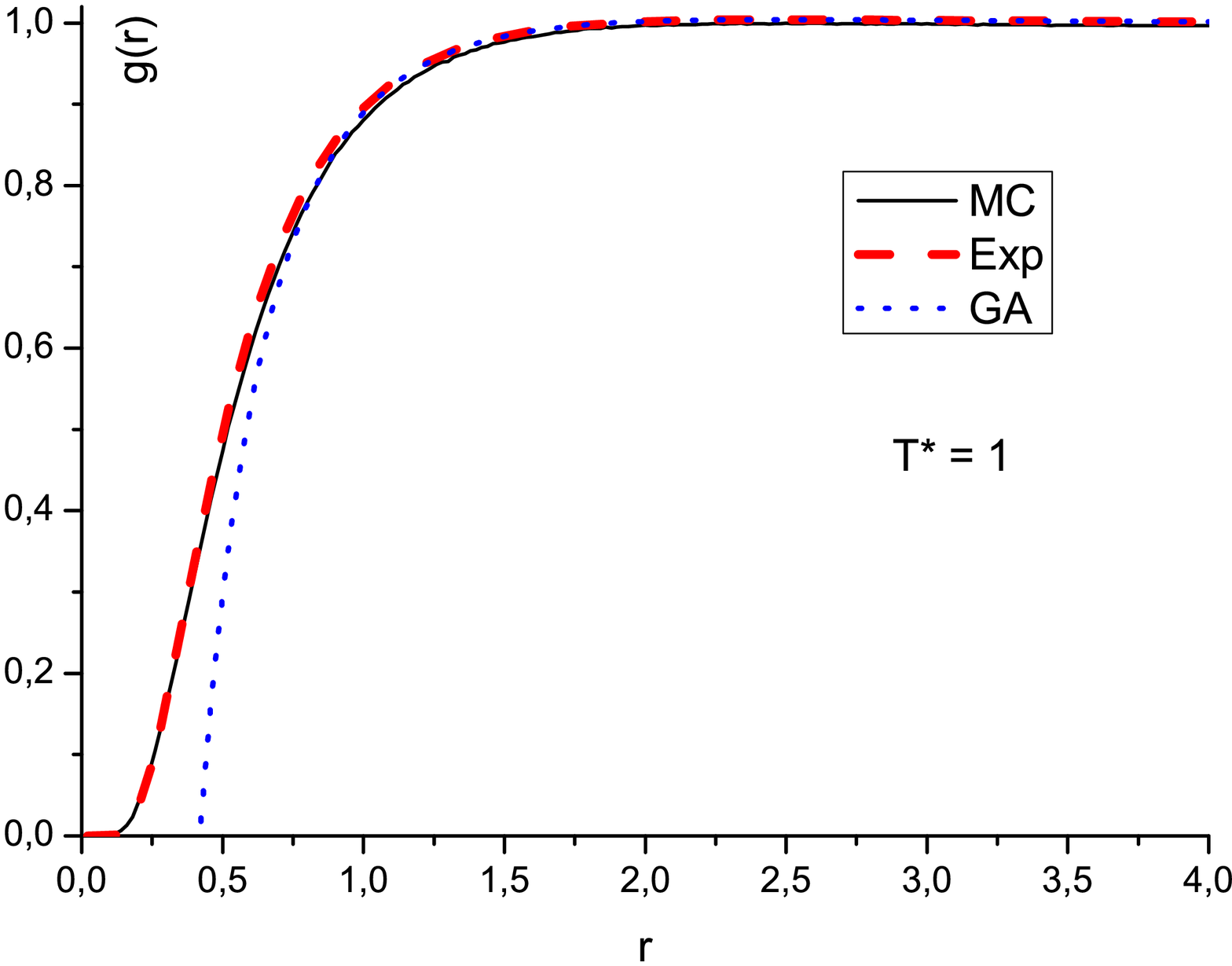}}
{\includegraphics[scale=0.24]{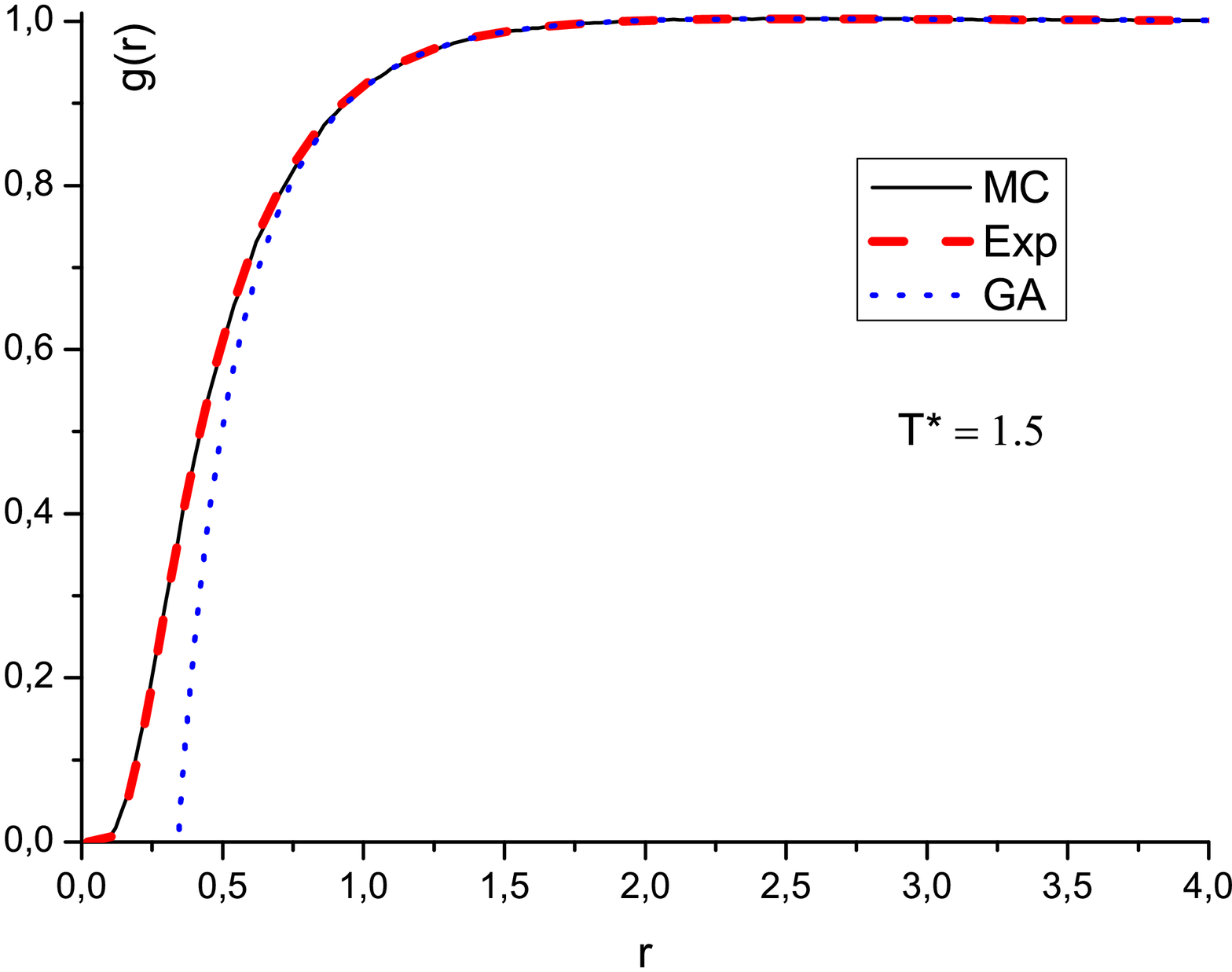}}
\caption{Analytical approximations (\ref{g}) and (\ref{g-exp}) and computer simulation results for the pair distribution function at different temperatures with $\omega =2,\tau =1.35, \rho^*=0.1$. "MC" corresponds to Monte Carlo simulations, "GA" is Gaussian approximation (\ref{g}) and "Exp" is exponential approximation (\ref{g-exp}).}
\label{fig2}
\end{figure}
The behavior of $g^b(r)$ at different temperatures and densities is illustrated in Fig. \ref{fig3}. We can see that with decreasing temperature the first peak of $g^b(r)$ increases and shifts to smaller distances. As the density increases at a fixed temperature the first peak of $g^b(r)$ decreases and shifts to smaller distances. Such a behavior is the result of softness of the model since it allows the particles to occupy the soft region as the density is increased and the temperature is decreased. We should also note that the agreement between theory and computer simulations results becomes worse as the temperature decreases. The height of the first peak increases faster in computer simulations as compared to theory.
 \begin{figure}
{\includegraphics[scale=0.3]{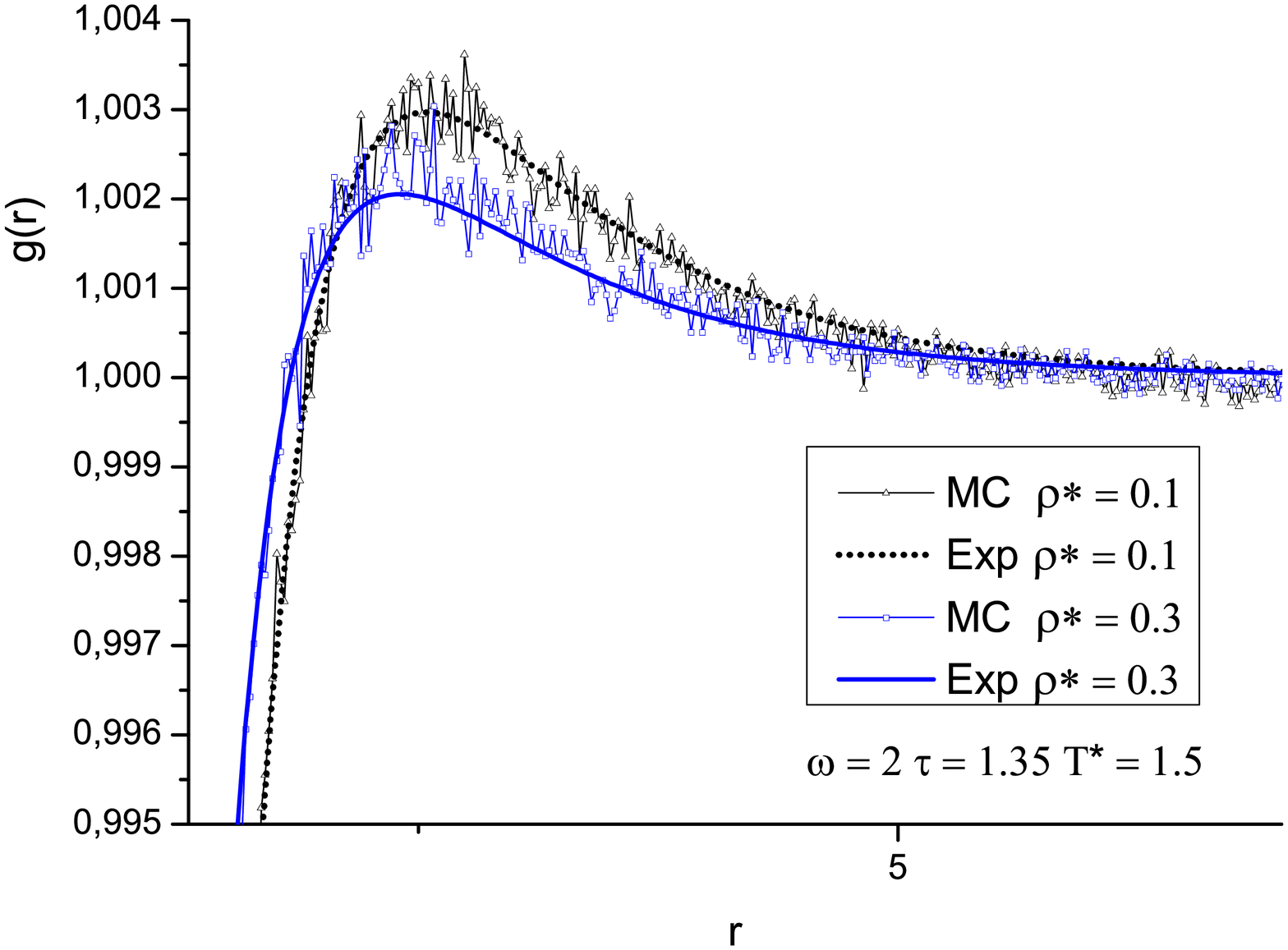}}
{\includegraphics[scale=0.3]{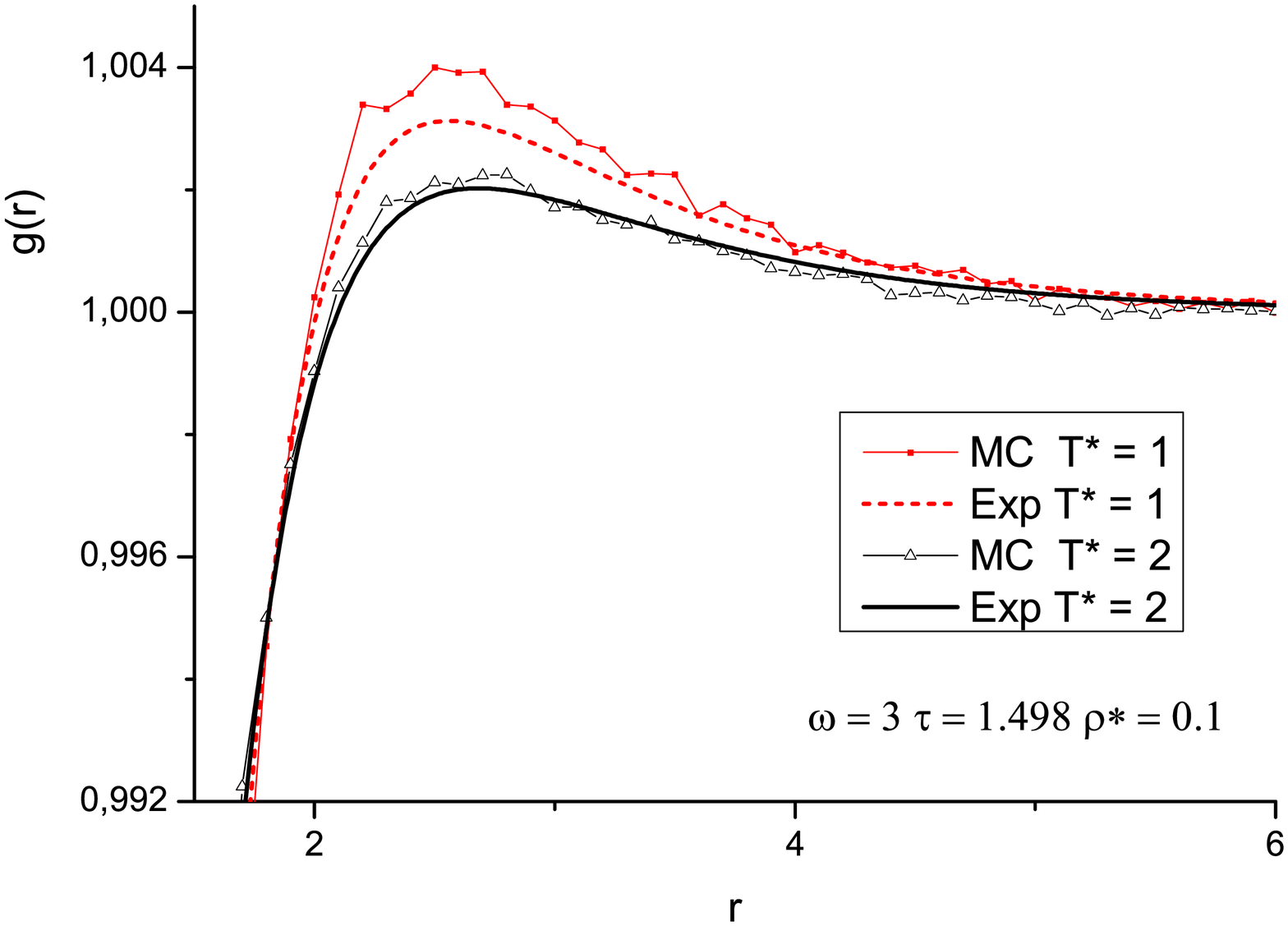}}
\caption{Analytical approximation (\ref{g-exp}) and computer simulations results for the pair distribution function at different densities (left) and different temperatures (right). The curves are labeled as in Fig. \ref{fig2}}
\label{fig3}
\end{figure}

\subsection{Density profile}
{\color{black}
According to (\ref{twoparts}) the density profile $\rho(z)$ can be presented as the sum of two parts:
$\rho^{MFA}(z)$ and $\rho^{fluct}(z)$. The first one is the result of the mean field approximation, which can be calculated
from the equation~(\ref{linearized_d}) derived from the linearized solution of the equation~(\ref{f3}) as it was done in \cite{preprint2013}. Another way to obtain $\rho^{MFA}(z)$ is to solve the equation~(\ref{f3}) numerically, which is obviously more precise.
To this aim we apply the Picard iterative method, where the numerical integrations are performed using the trapezoidal rule
with the step size $\Delta z=0.01\alpha_2$, while all needed integrations over $r$ are done analytically.
The cutoff for the two-Yukawa potential~(\ref{pot1}) is taken at the distance $r_c=12.0$,
at which the potential becomes negligibly small ($\nu(r_c)/\nu(r_{min})\sim 10^{-4}$, $r_{min}$ -- a position of the potential minimum).
Due to the hard wall presence from one side and the bulk phase from opposite side to the wall
the boundary conditions for $\rho^{MFA}(z)$ are defined as $\rho^{MFA}(z)=0$ if $z<0$ and $\rho^{MFA}(z)=\rho_b$ if $z>2 r_c$.
A precision of numerical solution for the density profile $\rho^{MFA}(z)$ is restricted by $\|\rho^{MFA}_{m+1}(z)-F[\rho^{MFA}_{m}(z)]\|<10^{-6}$, where $F[\rho^{MFA}_{m}(z)]$ is a right-hand side of the equation~(\ref{f3}) and $m$ is an iteration step.
It is worth mentioning that the equation~(\ref{f3}) is equivalent to the Euler-Lagrange equation, which
is usually used in the density functional theory within the mean field approximation for the fluid near a hard wall.

The density profiles, $\rho^{MFA}(z)$, obtained with a use of the linear approximation and the iterative method are presented in Fig.~\ref{fig4}.
As one can see in Fig.~\ref{fig4} the numerical solution of the equation~(\ref{f3}) can be interpolated very well by the linear approximation (\ref{linearized_d}). However, in the region of lower temperatures the linear approximation overestimates the density profile at intermediate distances. 
The contact values $\rho(0)$ calculated from MFA are essentially higher than those obtained from the simulations.
The general overestimation of MFA up to the first minimum of $\rho(z)$ is observed for all temperatures. 
A correction of $\rho^{MFA}(z)$ by the Gaussian fluctuation term $\rho^{fluct}(z)$ should improve the result.}

The Gaussian fluctuations term found from the solution of the inhomogeneous OZ equation with the Riemann boundary condition is given by the expression (\ref{density_profile}). It is observed in Fig. \ref{fig4} that the contribution from the fluctuations has a negative sign. This is an expected result since in \cite{molphys} it was shown that for a one-Yukawa fluid the fluctuation part of the density profile is negative for both attractive and repulsive interactions and produces the depletion effect. We have demonstrated that this term satisfies the contact theorem condition (\ref{CT}). Nevertheless, comparison with computer simulation results (Fig.\ref{fig4}) shows that this term leads to strong overestimation of the role of fluctuations. In addition, it gives a maximum of the profile at small distances from the wall which is not predicted by computer simulations. From Fig. \ref{fig4} one can also see that expression (\ref{density_profile}) strongly underestimates the contact value of the density profile. This is the consequence of the underestimated value of the pressure calculated from eq. (\ref{P}) which corresponds to approximation (\ref{g}) for the bulk PDF. Thus it would be more correct to calculate the pressure according to the virial theorem \cite{hansen}
\begin{figure}
{\includegraphics[scale=0.3]{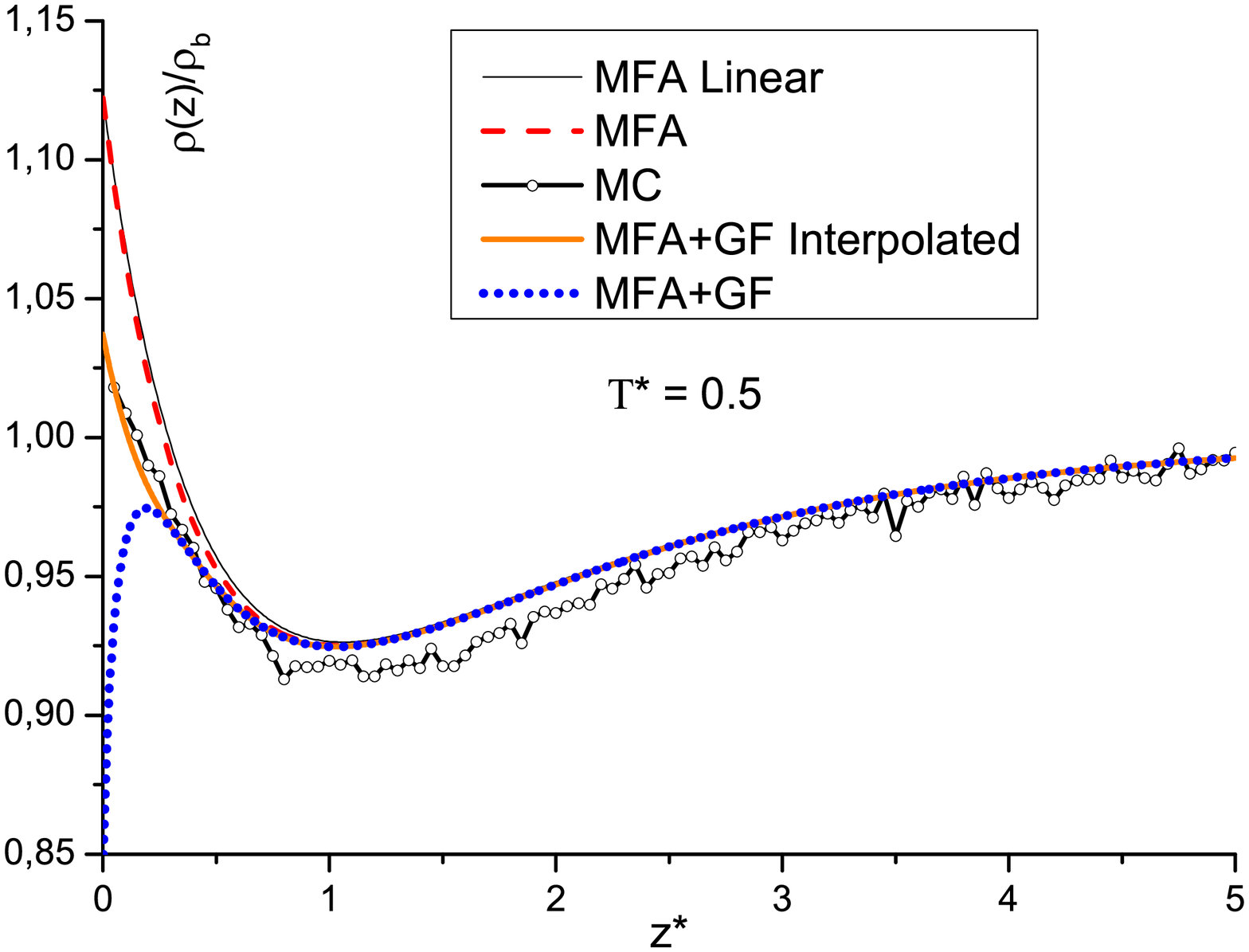}}
{\includegraphics[scale=0.3]{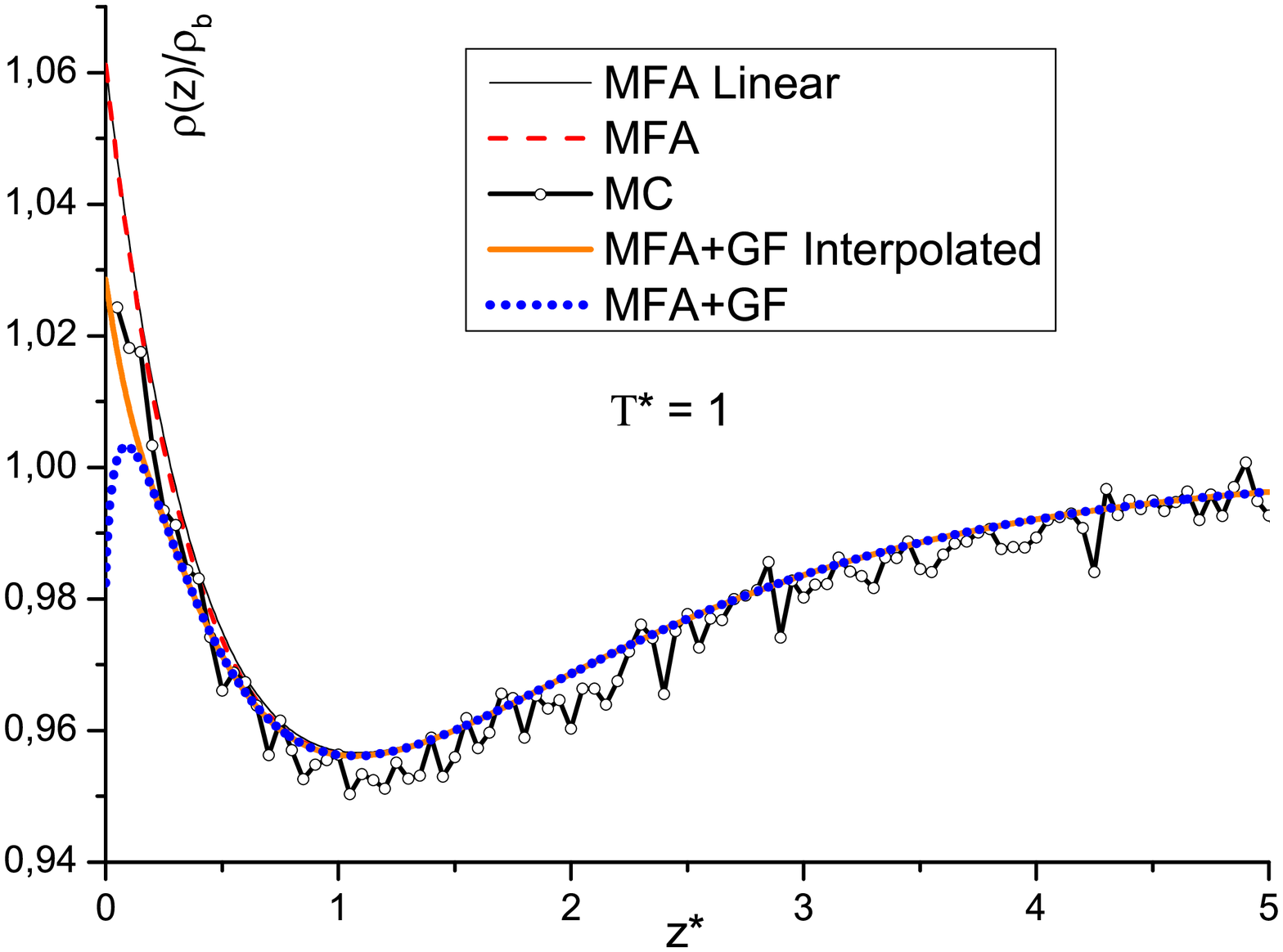}}
{\includegraphics[scale=0.3]{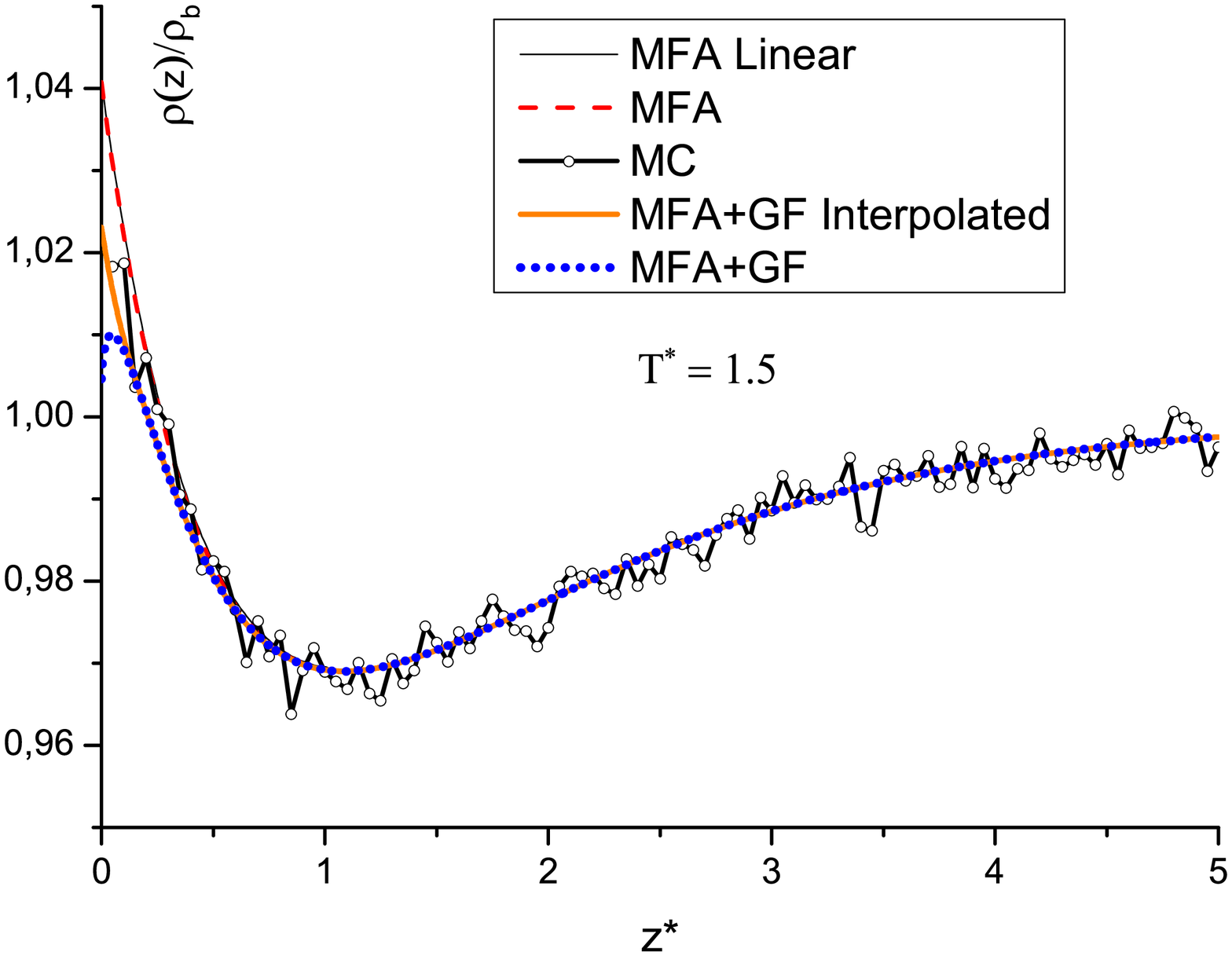}}
\caption{Density profile as a function of the reduced distance $z^*=z\alpha_2$. Label "MC" corresponds to Monte Carlo simulations, "MFA" is the numerical solution of eq. (\ref{f3}), "MFA Linear" is given by eq. (\ref{linearized_d}), "GF" is the Gaussian fluctuations term (\ref{density_profile}) and "MFA+GF Interpolated" corresponds to Gaussian approximation with interpolation (\ref{pade}) at small distances. Different graphs correspond to different temperatures at a fixed density $\rho^*=0.1$ with $\omega=2, \tau=1.35$.}
\label{fig4}
\end{figure}
\begin{align}
\frac{\beta P}{\rho _{b}}=1-\frac{2}{3}\pi\beta\int_{0}^{\infty}\frac{\partial\upsilon(r)}{\partial r}g^{b}(r)r^{3}dr.                                                 \label{virial}
\end{align}
and using the exponential approximation (\ref{g-exp}) for the bulk PDF.

Using this value of the pressure and the contact theorem we have corrected the density profile at small distances starting from point
$z_{0}$ which corresponds to the inflection point for the density fluctuation term (\ref{density_profile}). Interpolation of
$\rho(z)$ for the region of $z<z_{0}$ has revealed a rather accurate generalization in the form of Pad\'{e} approximant \cite{baker}
\begin{align}
\label{pade}
\rho(z)=\rho(0)/\left[1+Az+Bz^{2}+Cz^{3}\right]\qquad \text{for}\qquad z<z_{0},
\end{align}
where $\rho(0)$ is determined from the contact theorem and the pressure calculated from eq. (\ref{virial}). The constants $A,B,C$ are found from the continuity of $\rho(z)$ and its first derivative at the point $z=z_{0}$ and the fact that the second derivative equals zero at this point. This is the form we use to calculate the density profile. The results for the model with $\omega=2,\tau=1.35$ at density $\rho^*=0.1$ and different temperatures are shown in Fig.\ref{fig4}. One can see that the results of calculation are in good agreement with the computer simulations data.

In Fig. \ref{fig5} we present density profiles for a model with $\omega =3,\tau =1.498$ at different temperatures and densities. One can see that the contact value of the density increases and the minimum of the profile decreases as the temperature decreases. Below we will see that this can lead to non-trivial behavior of the adsorption as a function of the temperature. Likewise, the contact value of the density increases and the minimum of the profile decreases as the density increases at a fixed temperature. Due to the contact theorem and according to expression (\ref{virial}) the increase of the contact value of the DP with increasing density or decreasing temperature is connected with the respective increase of the fluid pressure in the bulk. The decrease of the minimum value of the DP with increasing density or decreasing temperature is defined mostly by the MFA. The fact that in the present model the mean field contribution is non-monotonous means that the fluid can have a layered-type structure which was not observed in the one-Yukawa case. The results obtained are in qualitative agreement with \cite{yu} where a hard core two-Yukawa fluid near a hard wall was studied by means of Monte-Carlo simulations and the density functional theory.
\begin{figure}
{\includegraphics[scale=0.3]{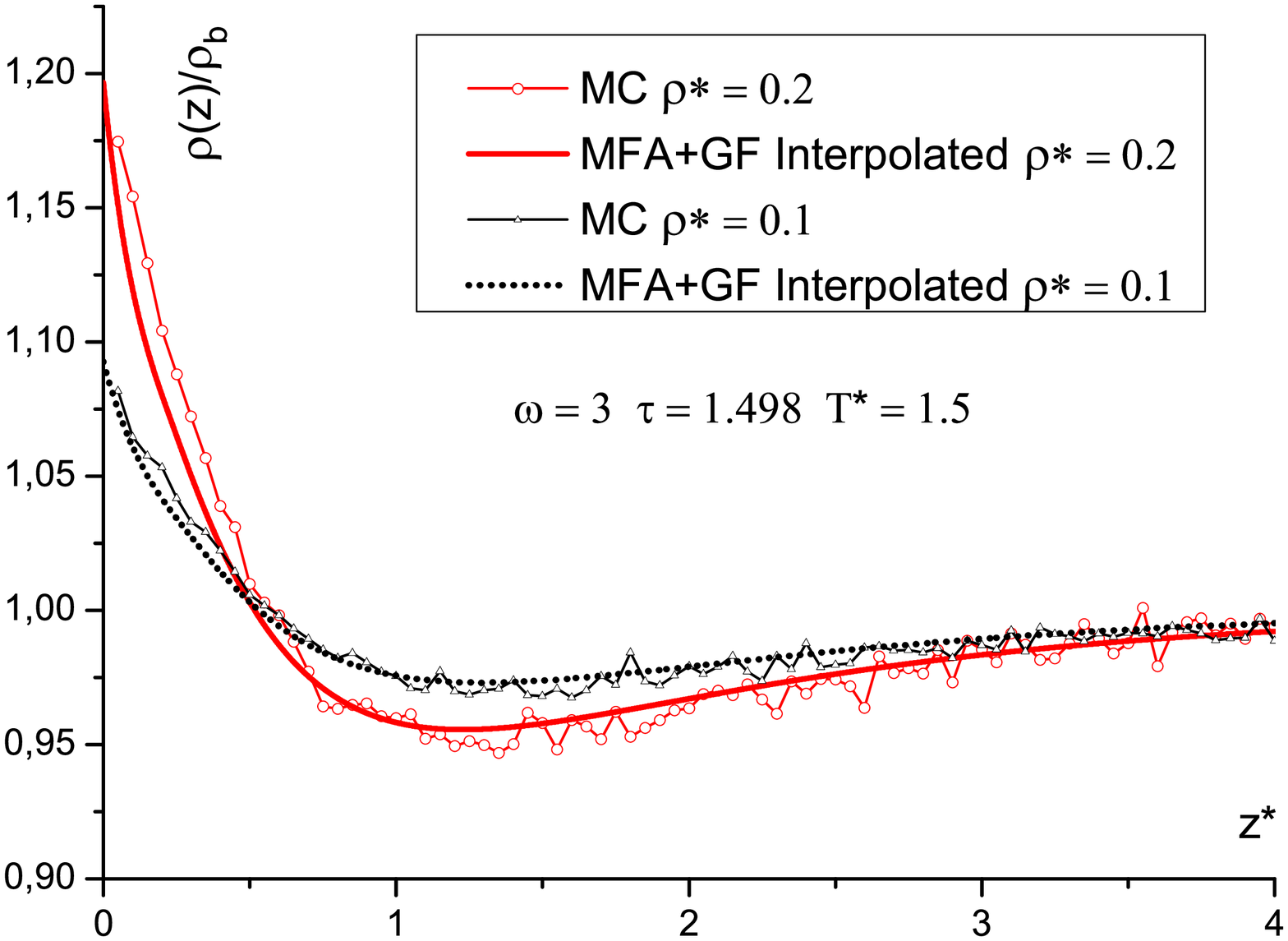}}
{\includegraphics[scale=0.3]{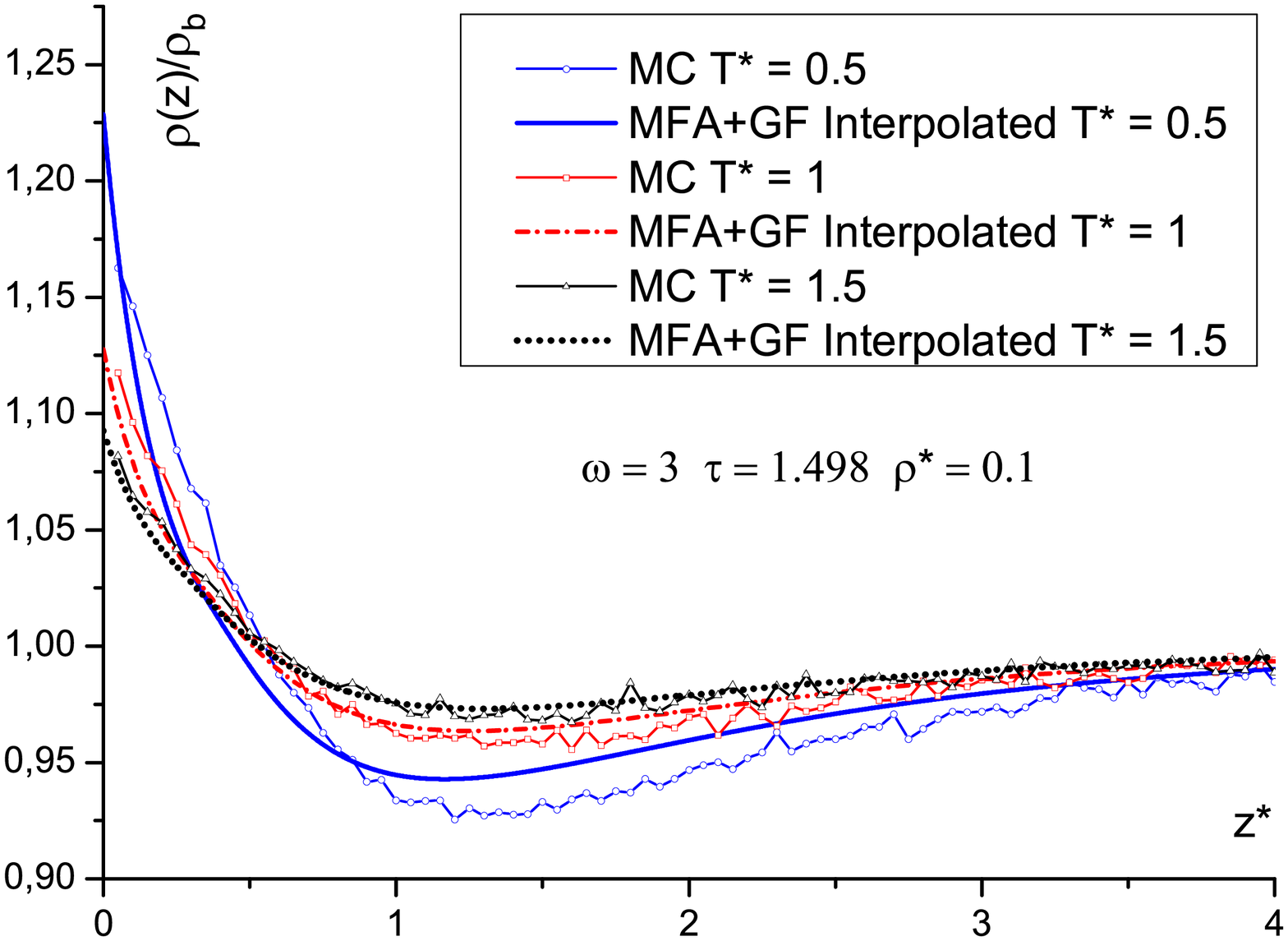}}
\caption{Theoretical approximations and computer simulations results for the density profile at different densities (left) and different temperatures (right). The curves are labeled as in Fig. \ref{fig4}}
\label{fig5}
\end{figure}

\subsection{Adsorption}

The adsorption coefficient (AC) defined by expr. (\ref{ads}) characterizes the excess of the density near the surface as compared to the bulk region. In accordance with (\ref{ads}) the AC can be presented as the sum of the mean field contribution $\Gamma_{MFA}$ and the fluctuation term $\Gamma_{fluct}$. As we have noted in \cite{preprint2013} the linearized MFA approximation can be positive or negative whereas the contribution of $\Gamma_{fluct}$ is negative for the approximation (\ref{gamma-f}). Unlike the mean field contribution, the contribution from
fluctuations $\Gamma_{fluct}$ is always negative. This result
is expected as in \cite{molphys} it was shown that for a
one-Yukawa fluid at a wall the fluctuation effects lead to
density depletion for both repulsive and attractive
interactions. In the region where $\Gamma_{MFA}$ is negative
the value of the total adsorption coefficient $\Gamma$ will be
negative. It is therefore more interesting to consider the
region in which $\Gamma_{MFA}$ is positive. In this case we
will have the competition between the MFA contribution and the
contribution
from fluctuations. In Fig. \ref{fig6} adsorption coefficients as functions of the temperature and the density for a model with $\omega=2.5,\tau=1.355$ are presented. For this case the mean field contribution $\Gamma_{MFA}$ is positive. As we can see from Fig. \ref{fig6} the linearized MFA overestimates the contribution to $\Gamma_{MFA}$ and the difference between the linearized and the non-linearized cases becomes more pronounced as the density increases or the temperature decreases. At high temperatures the role of the fluctuation term becomes negligible and the interpolated CA merges asymptotically with the MFA contribution. At moderate and lower temperatures, however, the Gaussian correction can modify significantly the MFA predictions. Notably, due to the competition between $\Gamma_{MFA}$ and $\Gamma_{fluct}$ the adsorption isotherm can display non-monotonous behavior as a function of the bulk density. Likewise, the adsorption isochore can be non-monotonous as a function of the temperature. Another interesting consequence of going beyond the MFA is that under certain conditions the CA can change sign as the temperature or the bulk density are varied. We should note that in \cite{yu} a similar effect was observed for a hard core two-Yukawa fluid in the framework of Monte-Carlo
simulations and the density functional theory.

\begin{figure}
{\includegraphics[scale=0.26]{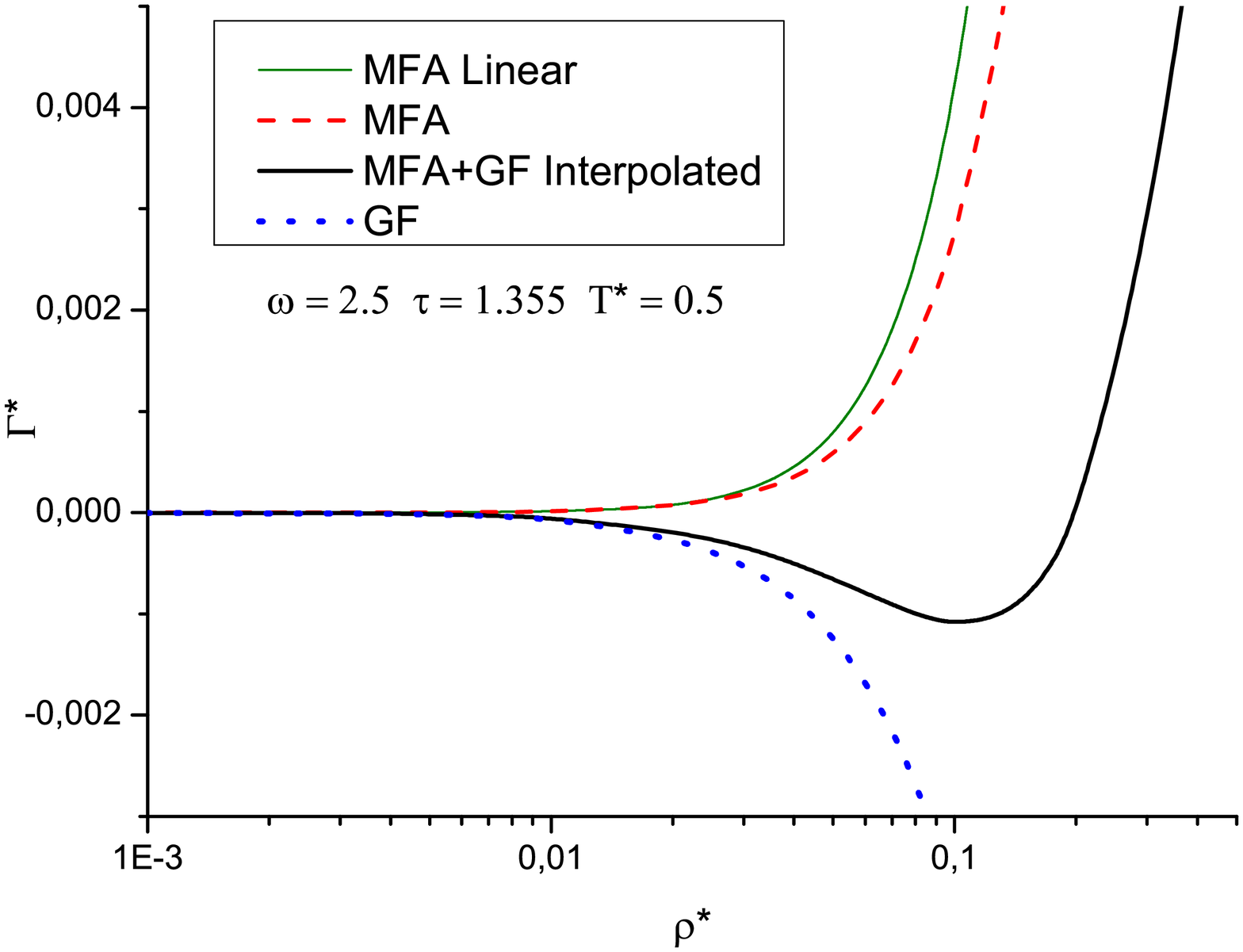}}
{\includegraphics[scale=0.26]{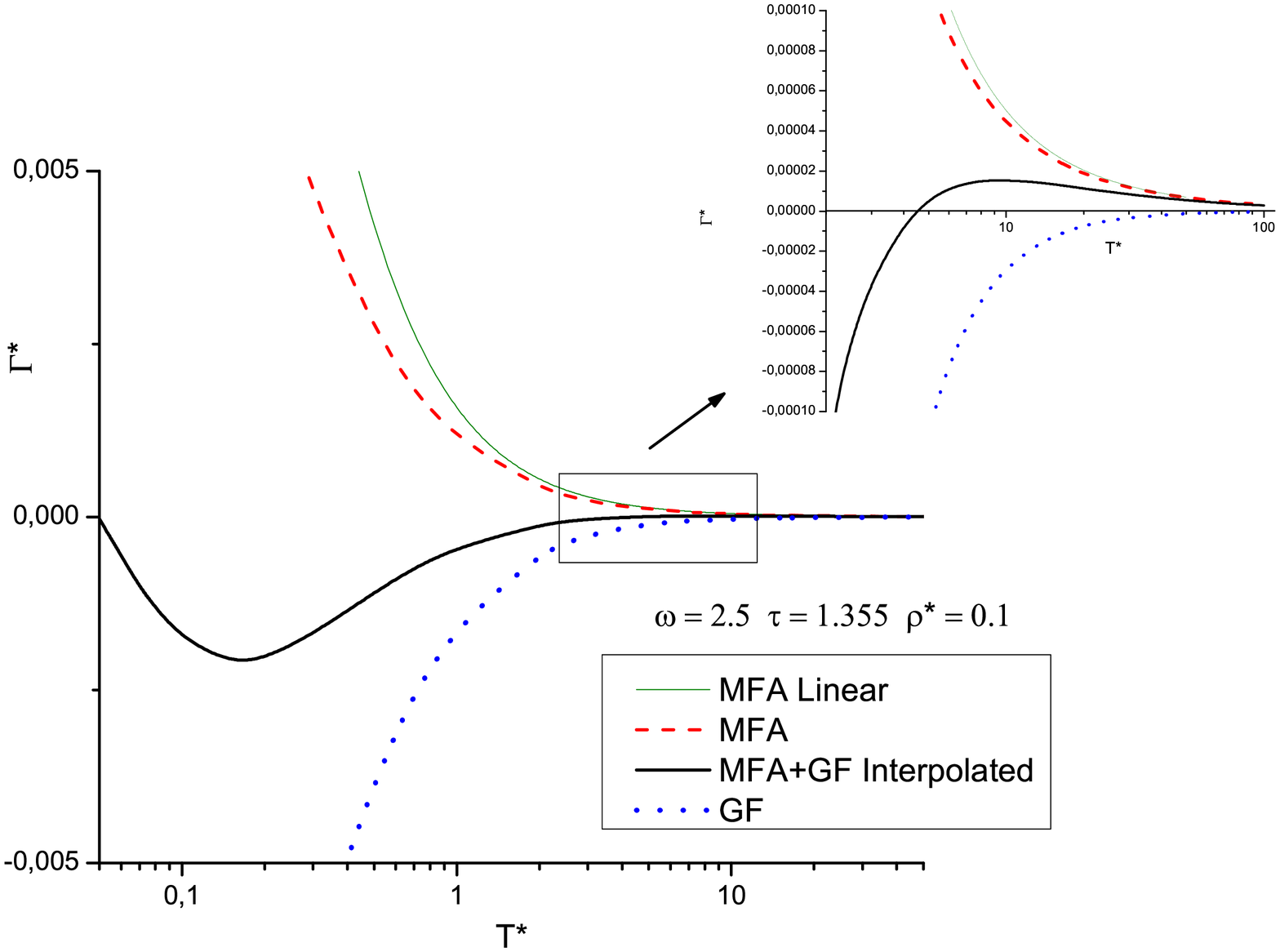}}
\caption{Reduced adsorption coefficient $\Gamma^*=\Gamma/\alpha_2^2$ as a function of the density at a fixed temperature (left) and as a function of the temperature at a fixed density (right).}
\label{fig6}
\end{figure}

\section{Conclusions}

In this work a field theoretical approach is applied to
describe a fluid interacting with a repulsive and an
attractive Yukawa potentials in the vicinity of a hard wall.
The results obtained are compared to a more simple one-Yukawa
model considered in our previous work \cite{molphys}. We derive
mean field equations that allow for a numerical evaluation of
the density profile. Subsequently the contact theorem is
validated employing a scheme that can by linearity be
generalized to a multi-Yukawa fluid. We find that unlike a one-Yukawa fluid, a
two-Yukawa fluid can have a non-monotonic profile even in the
mean field approximation. The linearized version of the profile
contains two generalized decays $\lambda_1$ and $\lambda_2$
which have a more complicated form than in the one-Yukawa case.
The results obtained in \cite{molphys} for an
attractive one-Yukawa case are not defined when
$\varkappa_2^2+\alpha_2^2<0$, that is for low temperatures,
high densities, or strongly attractive potentials. This
peculiarity is related to general problems in the description
of phase transitions in the framework of the Gaussian
fluctuations theory in the bulk. More specifically, it is the
so-called RPA-catastrophe which is caused by an incorrect
treatment of short-range correlations and can be removed by
including the repulsive interactions \cite{wheeler}. Compared
to an attractive one-Yukawa case we thus show that
generalization of the interaction potential to the sum of a
repulsive and an attractive parts makes the profile decays well
defined for all temperatures and densities.

Beyond the mean field approximation we study the impact of
Gaussian fluctuations on thermodynamic and structural
properties of the fluid. Analytical expressions for the free
energy, the pressure, the chemical potential, and the
correlation function are derived. Subsequently we find a
correction to the density profile due to fluctuations and show
that fluctuations always lead to depletion. We show analytically that the
fluctuation terms of the pressure and of the density contact
value satisfy the contact theorem. However, comparison with the computer simulations data has revealed that
the contribution from fluctuations leads to strong overestimation of the role of fluctuations. It produces a maximum of the profile at small distances to the wall which is not predicted by computer simulations. The fluctuation term also strongly underestimates the contact value of the density profile. In accordance with the contact theorem this phenomenon is the results of the incorrect prediction of the bulk pressure in the framework of the Gaussian approximation. We also show that the Gaussian approximation leads to incorrect behavior of the bulk pair distribution function at small interparticle distances. In order to improve the bulk pair distribution function at small distances we propose an exponential approximation which ensures the correct behavior of the PDF at small distances and reproduces the prediction of the Gaussian approximation at larger distances. The exponential form of the PDF also agrees very well with the computer simulations results. The pressure calculated from the exponential form of the PDF ensures the correct contact value of the density profile. We use this result to improve the description of the density profile at small distances to the wall. The results calculated via such an interpolation procedure are in a very good agreement with the computer simulations data.

Next we study the adsorption
coefficient and its dependence on the bulk density and the
temperature. Unlike the mean field
part, the contribution from fluctuations is always negative. We
consider the case when there is a competition between the two
contributions. It is found that at higher temperatures the mean
field term dominates, but as the temperature decreases the
fluctuation effects become increasingly more important. As a
result, non-monotonic adsorption curves are found for some
systems. The behaviors of the density profile and of the
adsorption isotherm described in this paper are in qualitative
agreement with the results of \cite{yu}, where a hard core
two-Yukawa fluid was studied by means of Monte-Carlo
simulations and the density functional theory.

\begin{acknowledgments}
The authors are grateful
for the support of the National Academy of Sciences of Ukraine
and the Centre National de la Recherche Scientifique
(CNRS) in the framework of the PICS project.
\end{acknowledgments}

\bibliography{paper}

\begin{thebibliography}{36}%
\makeatletter
\providecommand \@ifxundefined [1]{%
 \@ifx{#1\undefined}
}%
\providecommand \@ifnum [1]{%
 \ifnum #1\expandafter \@firstoftwo
 \else \expandafter \@secondoftwo
 \fi
}%
\providecommand \@ifx [1]{%
 \ifx #1\expandafter \@firstoftwo
 \else \expandafter \@secondoftwo
 \fi
}%
\providecommand \natexlab [1]{#1}%
\providecommand \enquote  [1]{``#1''}%
\providecommand \bibnamefont  [1]{#1}%
\providecommand \bibfnamefont [1]{#1}%
\providecommand \citenamefont [1]{#1}%
\providecommand \href@noop [0]{\@secondoftwo}%
\providecommand \href [0]{\begingroup \@sanitize@url \@href}%
\providecommand \@href[1]{\@@startlink{#1}\@@href}%
\providecommand \@@href[1]{\endgroup#1\@@endlink}%
\providecommand \@sanitize@url [0]{\catcode `\\12\catcode `\$12\catcode
  `\&12\catcode `\#12\catcode `\^12\catcode `\_12\catcode `\%12\relax}%
\providecommand \@@startlink[1]{}%
\providecommand \@@endlink[0]{}%
\providecommand \url  [0]{\begingroup\@sanitize@url \@url }%
\providecommand \@url [1]{\endgroup\@href {#1}{\urlprefix }}%
\providecommand \urlprefix  [0]{URL }%
\providecommand \Eprint [0]{\href }%
\providecommand \doibase [0]{http://dx.doi.org/}%
\providecommand \selectlanguage [0]{\@gobble}%
\providecommand \bibinfo  [0]{\@secondoftwo}%
\providecommand \bibfield  [0]{\@secondoftwo}%
\providecommand \translation [1]{[#1]}%
\providecommand \BibitemOpen [0]{}%
\providecommand \bibitemStop [0]{}%
\providecommand \bibitemNoStop [0]{.\EOS\space}%
\providecommand \EOS [0]{\spacefactor3000\relax}%
\providecommand \BibitemShut  [1]{\csname bibitem#1\endcsname}%
\let\auto@bib@innerbib\@empty
\bibitem [{\citenamefont {{Kalyuzhnyi}}\ and\ \citenamefont
  {{Cummings}}(1996)}]{yukal1996}%
  \BibitemOpen
  \bibfield  {author} {\bibinfo {author} {\bibfnamefont {Y.}~\bibnamefont
  {{Kalyuzhnyi}}}\ and\ \bibinfo {author} {\bibfnamefont {P.}~\bibnamefont
  {{Cummings}}},\ }\href@noop {} {\bibfield  {journal} {\bibinfo  {journal}
  {Mol. Phys.}\ }\textbf {\bibinfo {volume} {87}},\ \bibinfo {pages} {1459}
  (\bibinfo {year} {1996})}\BibitemShut {NoStop}%
\bibitem [{\citenamefont {{Tang}}, \citenamefont {{Tong}},\ and\ \citenamefont
  {{Lu}}(1997)}]{tang}%
  \BibitemOpen
  \bibfield  {author} {\bibinfo {author} {\bibfnamefont {Y.}~\bibnamefont
  {{Tang}}}, \bibinfo {author} {\bibfnamefont {Z.}~\bibnamefont {{Tong}}}, \
  and\ \bibinfo {author} {\bibfnamefont {B.-Y.}\ \bibnamefont {{Lu}}},\
  }\href@noop {} {\bibfield  {journal} {\bibinfo  {journal} {Fluid Phase
  Equilibr.}\ }\textbf {\bibinfo {volume} {134}},\ \bibinfo {pages} {21}
  (\bibinfo {year} {1997})}\BibitemShut {NoStop}%
\bibitem [{\citenamefont {Wu}\ and\ \citenamefont {Gao}(2005)}]{wu-gao}%
  \BibitemOpen
  \bibfield  {author} {\bibinfo {author} {\bibfnamefont {J.}~\bibnamefont
  {Wu}}\ and\ \bibinfo {author} {\bibfnamefont {J.}~\bibnamefont {Gao}},\
  }\href@noop {} {\bibfield  {journal} {\bibinfo  {journal} {J. Phys. Chem. B}\
  }\textbf {\bibinfo {volume} {109}},\ \bibinfo {pages} {21342} (\bibinfo
  {year} {2005})}\BibitemShut {NoStop}%
\bibitem [{\citenamefont {{Lin}}, \citenamefont {{Li}},\ and\ \citenamefont
  {{Lu}}(2001)}]{lin}%
  \BibitemOpen
  \bibfield  {author} {\bibinfo {author} {\bibfnamefont {Y.-Z.}\ \bibnamefont
  {{Lin}}}, \bibinfo {author} {\bibfnamefont {Y.-G.}\ \bibnamefont {{Li}}}, \
  and\ \bibinfo {author} {\bibfnamefont {J.-F.}\ \bibnamefont {{Lu}}},\
  }\href@noop {} {\bibfield  {journal} {\bibinfo  {journal} {J. Colloidal
  Interface Sci.}\ }\textbf {\bibinfo {volume} {239}},\ \bibinfo {pages} {58}
  (\bibinfo {year} {2001})}\BibitemShut {NoStop}%
\bibitem [{\citenamefont {{Archer}}\ and\ \citenamefont
  {{Evans}}(2007)}]{archer}%
  \BibitemOpen
  \bibfield  {author} {\bibinfo {author} {\bibfnamefont {A.~J.}\ \bibnamefont
  {{Archer}}}\ and\ \bibinfo {author} {\bibfnamefont {R.}~\bibnamefont
  {{Evans}}},\ }\href@noop {} {\bibfield  {journal} {\bibinfo  {journal} {J.
  Chem. Phys.}\ }\textbf {\bibinfo {volume} {126}},\ \bibinfo {pages} {014104}
  (\bibinfo {year} {2007})}\BibitemShut {NoStop}%
\bibitem [{\citenamefont {Archer}\ \emph {et~al.}(2007)\citenamefont {Archer},
  \citenamefont {Pini}, \citenamefont {Evans},\ and\ \citenamefont
  {Reatto}}]{archer_pini}%
  \BibitemOpen
  \bibfield  {author} {\bibinfo {author} {\bibfnamefont {A.}~\bibnamefont
  {Archer}}, \bibinfo {author} {\bibfnamefont {D.}~\bibnamefont {Pini}},
  \bibinfo {author} {\bibfnamefont {R.}~\bibnamefont {Evans}}, \ and\ \bibinfo
  {author} {\bibfnamefont {L.}~\bibnamefont {Reatto}},\ }\href@noop {}
  {\bibfield  {journal} {\bibinfo  {journal} {J. Chem. Phys.}\ }\textbf
  {\bibinfo {volume} {126}},\ \bibinfo {pages} {014104} (\bibinfo {year}
  {2007})}\BibitemShut {NoStop}%
\bibitem [{\citenamefont {{Lin}}, \citenamefont {{Chen}},\ and\ \citenamefont
  {{Chen}}(2005)}]{lin_chen}%
  \BibitemOpen
  \bibfield  {author} {\bibinfo {author} {\bibfnamefont {Y.}~\bibnamefont
  {{Lin}}}, \bibinfo {author} {\bibfnamefont {W.-R.}\ \bibnamefont {{Chen}}}, \
  and\ \bibinfo {author} {\bibfnamefont {S.-H.}\ \bibnamefont {{Chen}}},\
  }\href@noop {} {\bibfield  {journal} {\bibinfo  {journal} {J. Phys. Chem. B}\
  }\textbf {\bibinfo {volume} {122}},\ \bibinfo {pages} {044507} (\bibinfo
  {year} {2005})}\BibitemShut {NoStop}%
\bibitem [{\citenamefont {{Kalyuzhnyi}}\ \emph {et~al.}(2004)\citenamefont
  {{Kalyuzhnyi}}, \citenamefont {{McCabe}}, \citenamefont {{Whitebay}},\ and\
  \citenamefont {{Cummings}}}]{kalyuzhnyi2}%
  \BibitemOpen
  \bibfield  {author} {\bibinfo {author} {\bibfnamefont {Y.}~\bibnamefont
  {{Kalyuzhnyi}}}, \bibinfo {author} {\bibfnamefont {C.}~\bibnamefont
  {{McCabe}}}, \bibinfo {author} {\bibfnamefont {E.}~\bibnamefont
  {{Whitebay}}}, \ and\ \bibinfo {author} {\bibfnamefont {P.}~\bibnamefont
  {{Cummings}}},\ }\href@noop {} {\bibfield  {journal} {\bibinfo  {journal} {J.
  Chem. Phys.}\ }\textbf {\bibinfo {volume} {121}},\ \bibinfo {pages} {8128}
  (\bibinfo {year} {2004})}\BibitemShut {NoStop}%
\bibitem [{\citenamefont {{Holovko}}\ and\ \citenamefont
  {{Sokolovska}}(1999)}]{holovko1999}%
  \BibitemOpen
  \bibfield  {author} {\bibinfo {author} {\bibfnamefont {M.}~\bibnamefont
  {{Holovko}}}\ and\ \bibinfo {author} {\bibfnamefont {T.}~\bibnamefont
  {{Sokolovska}}},\ }\href@noop {} {\bibfield  {journal} {\bibinfo  {journal}
  {J. Mol. Liq.}\ }\textbf {\bibinfo {volume} {82}},\ \bibinfo {pages} {161}
  (\bibinfo {year} {1999})}\BibitemShut {NoStop}%
\bibitem [{\citenamefont {Kravtsiv}, \citenamefont {{Holovko}},\ and\
  \citenamefont {{Di Caprio}}(2013)}]{kravtsiv2013}%
  \BibitemOpen
  \bibfield  {author} {\bibinfo {author} {\bibfnamefont {I.}~\bibnamefont
  {Kravtsiv}}, \bibinfo {author} {\bibfnamefont {M.}~\bibnamefont {{Holovko}}},
  \ and\ \bibinfo {author} {\bibfnamefont {D.}~\bibnamefont {{Di Caprio}}},\
  }\href@noop {} {\bibfield  {journal} {\bibinfo  {journal} {Mol. Phys.}\
  }\textbf {\bibinfo {volume} {111}},\ \bibinfo {pages} {1023} (\bibinfo {year}
  {2013})}\BibitemShut {NoStop}%
\bibitem [{\citenamefont {{Waisman}}(1973)}]{Waisman}%
  \BibitemOpen
  \bibfield  {author} {\bibinfo {author} {\bibfnamefont {E.}~\bibnamefont
  {{Waisman}}},\ }\href@noop {} {\bibfield  {journal} {\bibinfo  {journal}
  {Mol. Phys.}\ }\textbf {\bibinfo {volume} {25}},\ \bibinfo {pages} {45}
  (\bibinfo {year} {1973})}\BibitemShut {NoStop}%
\bibitem [{\citenamefont {Ginosa}(1986)}]{ginosa}%
  \BibitemOpen
  \bibfield  {author} {\bibinfo {author} {\bibfnamefont {M.}~\bibnamefont
  {Ginosa}},\ }\href@noop {} {\bibfield  {journal} {\bibinfo  {journal} {J.
  Phys. Soc. Japan}\ }\textbf {\bibinfo {volume} {55}},\ \bibinfo {pages} {95}
  (\bibinfo {year} {1986})}\BibitemShut {NoStop}%
\bibitem [{\citenamefont {{Hoye}}\ and\ \citenamefont {{Blum}}(1978)}]{hoye}%
  \BibitemOpen
  \bibfield  {author} {\bibinfo {author} {\bibfnamefont {J.}~\bibnamefont
  {{Hoye}}}\ and\ \bibinfo {author} {\bibfnamefont {L.}~\bibnamefont
  {{Blum}}},\ }\href@noop {} {\bibfield  {journal} {\bibinfo  {journal} {J.
  Stat. Phys.}\ }\textbf {\bibinfo {volume} {19}},\ \bibinfo {pages} {317}
  (\bibinfo {year} {1978})}\BibitemShut {NoStop}%
\bibitem [{\citenamefont {{Lin}}, \citenamefont {{Li}},\ and\ \citenamefont
  {{Lu}}(2004)}]{lin-li}%
  \BibitemOpen
  \bibfield  {author} {\bibinfo {author} {\bibfnamefont {Y.}~\bibnamefont
  {{Lin}}}, \bibinfo {author} {\bibfnamefont {Y.-G.}\ \bibnamefont {{Li}}}, \
  and\ \bibinfo {author} {\bibfnamefont {L.-F.}\ \bibnamefont {{Lu}}},\
  }\href@noop {} {\bibfield  {journal} {\bibinfo  {journal} {Mol. Phys.}\
  }\textbf {\bibinfo {volume} {102}},\ \bibinfo {pages} {63} (\bibinfo {year}
  {2004})}\BibitemShut {NoStop}%
\bibitem [{\citenamefont {Likos}\ \emph {et~al.}(1998)\citenamefont {Likos},
  \citenamefont {L\"{o}wen}, \citenamefont {Watzlawek}, \citenamefont {Ablas},
  \citenamefont {Jucknischke}, \citenamefont {Algaier},\ and\ \citenamefont
  {Richter}}]{likos}%
  \BibitemOpen
  \bibfield  {author} {\bibinfo {author} {\bibfnamefont {C.}~\bibnamefont
  {Likos}}, \bibinfo {author} {\bibfnamefont {H.}~\bibnamefont {L\"{o}wen}},
  \bibinfo {author} {\bibfnamefont {M.}~\bibnamefont {Watzlawek}}, \bibinfo
  {author} {\bibfnamefont {B.}~\bibnamefont {Ablas}}, \bibinfo {author}
  {\bibfnamefont {O.}~\bibnamefont {Jucknischke}}, \bibinfo {author}
  {\bibfnamefont {J.}~\bibnamefont {Algaier}}, \ and\ \bibinfo {author}
  {\bibfnamefont {D.}~\bibnamefont {Richter}},\ }\href@noop {} {\bibfield
  {journal} {\bibinfo  {journal} {Phys. Rev. Lett.}\ }\textbf {\bibinfo
  {volume} {80}},\ \bibinfo {pages} {4450} (\bibinfo {year}
  {1998})}\BibitemShut {NoStop}%
\bibitem [{\citenamefont {Camargo}\ and\ \citenamefont
  {Likos}(2009)}]{camargo}%
  \BibitemOpen
  \bibfield  {author} {\bibinfo {author} {\bibfnamefont {M.}~\bibnamefont
  {Camargo}}\ and\ \bibinfo {author} {\bibfnamefont {C.}~\bibnamefont
  {Likos}},\ }\href@noop {} {\bibfield  {journal} {\bibinfo  {journal} {J.
  Chem. Phys.}\ }\textbf {\bibinfo {volume} {134}},\ \bibinfo {pages} {204904}
  (\bibinfo {year} {2009})}\BibitemShut {NoStop}%
\bibitem [{\citenamefont {{Holovko}}, \citenamefont {{Kravtsiv}},\ and\
  \citenamefont {{Soviak}}(2009)}]{soviak}%
  \BibitemOpen
  \bibfield  {author} {\bibinfo {author} {\bibfnamefont {M.}~\bibnamefont
  {{Holovko}}}, \bibinfo {author} {\bibfnamefont {I.}~\bibnamefont
  {{Kravtsiv}}}, \ and\ \bibinfo {author} {\bibfnamefont {{\relax
  E}.}~\bibnamefont {{Soviak}}},\ }\href@noop {} {\bibfield  {journal}
  {\bibinfo  {journal} {Condens. Matter Phys.}\ }\textbf {\bibinfo {volume}
  {12}},\ \bibinfo {pages} {137} (\bibinfo {year} {2009})}\BibitemShut
  {NoStop}%
\bibitem [{\citenamefont {{Di Caprio}}\ \emph {et~al.}(2011)\citenamefont {{Di
  Caprio}}, \citenamefont {{Stafiej}}, \citenamefont {{Holovko}},\ and\
  \citenamefont {{Kravtsiv}}}]{molphys}%
  \BibitemOpen
  \bibfield  {author} {\bibinfo {author} {\bibfnamefont {D.}~\bibnamefont {{Di
  Caprio}}}, \bibinfo {author} {\bibfnamefont {J.}~\bibnamefont {{Stafiej}}},
  \bibinfo {author} {\bibfnamefont {M.}~\bibnamefont {{Holovko}}}, \ and\
  \bibinfo {author} {\bibfnamefont {I.}~\bibnamefont {{Kravtsiv}}},\
  }\href@noop {} {\bibfield  {journal} {\bibinfo  {journal} {Mol. Phys.}\
  }\textbf {\bibinfo {volume} {109}},\ \bibinfo {pages} {695} (\bibinfo {year}
  {2011})}\BibitemShut {NoStop}%
\bibitem [{\citenamefont {Olivares-Rivas}\ \emph {et~al.}(1997)\citenamefont
  {Olivares-Rivas}, \citenamefont {Degreve}, \citenamefont {Henderson},\ and\
  \citenamefont {Quintana}}]{olivares}%
  \BibitemOpen
  \bibfield  {author} {\bibinfo {author} {\bibfnamefont {W.}~\bibnamefont
  {Olivares-Rivas}}, \bibinfo {author} {\bibfnamefont {L.}~\bibnamefont
  {Degreve}}, \bibinfo {author} {\bibfnamefont {D.}~\bibnamefont {Henderson}},
  \ and\ \bibinfo {author} {\bibfnamefont {J.}~\bibnamefont {Quintana}},\
  }\href@noop {} {\bibfield  {journal} {\bibinfo  {journal} {J. Chem. Phys.}\
  }\textbf {\bibinfo {volume} {107}},\ \bibinfo {pages} {8147} (\bibinfo {year}
  {1997})}\BibitemShut {NoStop}%
\bibitem [{\citenamefont {You}, \citenamefont {Yu},\ and\ \citenamefont
  {Gao}(2005)}]{you}%
  \BibitemOpen
  \bibfield  {author} {\bibinfo {author} {\bibfnamefont {F.}~\bibnamefont
  {You}}, \bibinfo {author} {\bibfnamefont {Y.}~\bibnamefont {Yu}}, \ and\
  \bibinfo {author} {\bibfnamefont {G.}~\bibnamefont {Gao}},\ }\href@noop {}
  {\bibfield  {journal} {\bibinfo  {journal} {J. Phys. Chem. B}\ }\textbf
  {\bibinfo {volume} {109}},\ \bibinfo {pages} {3512} (\bibinfo {year}
  {2005})}\BibitemShut {NoStop}%
\bibitem [{\citenamefont {Tang}\ and\ \citenamefont {Wu}(2004)}]{tang-wu}%
  \BibitemOpen
  \bibfield  {author} {\bibinfo {author} {\bibfnamefont {Y.}~\bibnamefont
  {Tang}}\ and\ \bibinfo {author} {\bibfnamefont {J.}~\bibnamefont {Wu}},\
  }\href@noop {} {\bibfield  {journal} {\bibinfo  {journal} {Phys. Rev. E}\
  }\textbf {\bibinfo {volume} {70}},\ \bibinfo {pages} {011201} (\bibinfo
  {year} {2004})}\BibitemShut {NoStop}%
\bibitem [{\citenamefont {{Yu}}\ \emph {et~al.}(2006)\citenamefont {{Yu}},
  \citenamefont {{You}}, \citenamefont {{Tang}}, \citenamefont {{Gao}},\ and\
  \citenamefont {{Li}}}]{yu}%
  \BibitemOpen
  \bibfield  {author} {\bibinfo {author} {\bibfnamefont {Y.}~\bibnamefont
  {{Yu}}}, \bibinfo {author} {\bibfnamefont {F.}~\bibnamefont {{You}}},
  \bibinfo {author} {\bibfnamefont {Y.}~\bibnamefont {{Tang}}}, \bibinfo
  {author} {\bibfnamefont {G.}~\bibnamefont {{Gao}}}, \ and\ \bibinfo {author}
  {\bibfnamefont {Y.}~\bibnamefont {{Li}}},\ }\href@noop {} {\bibfield
  {journal} {\bibinfo  {journal} {J. Phys. Chem. B}\ }\textbf {\bibinfo
  {volume} {110}},\ \bibinfo {pages} {334} (\bibinfo {year}
  {2006})}\BibitemShut {NoStop}%
\bibitem [{\citenamefont {{Kim}}\ and\ \citenamefont {{Kim}}(2012)}]{kim_kim}%
  \BibitemOpen
  \bibfield  {author} {\bibinfo {author} {\bibfnamefont {E.-Y.}\ \bibnamefont
  {{Kim}}}\ and\ \bibinfo {author} {\bibfnamefont {S.-C.}\ \bibnamefont
  {{Kim}}},\ }\href@noop {} {\bibfield  {journal} {\bibinfo  {journal} {Phys.
  Rev. E}\ }\textbf {\bibinfo {volume} {85}},\ \bibinfo {pages} {051203}
  (\bibinfo {year} {2012})}\BibitemShut {NoStop}%
\bibitem [{\citenamefont {Henderson}, \citenamefont {Blum},\ and\ \citenamefont
  {Lebowitz}(1979)}]{HendersonBlumLebowitz1979}%
  \BibitemOpen
  \bibfield  {author} {\bibinfo {author} {\bibfnamefont {D.}~\bibnamefont
  {Henderson}}, \bibinfo {author} {\bibfnamefont {L.}~\bibnamefont {Blum}}, \
  and\ \bibinfo {author} {\bibfnamefont {J.}~\bibnamefont {Lebowitz}},\
  }\href@noop {} {\bibfield  {journal} {\bibinfo  {journal} {J. Electroanal.
  Phys.}\ }\textbf {\bibinfo {volume} {102}},\ \bibinfo {pages} {315} (\bibinfo
  {year} {1979})}\BibitemShut {NoStop}%
\bibitem [{\citenamefont {Holovko}, \citenamefont {Badiali},\ and\
  \citenamefont {di~Caprio}(2005)}]{HolovkoBadialidiCaprio2005}%
  \BibitemOpen
  \bibfield  {author} {\bibinfo {author} {\bibfnamefont {M.}~\bibnamefont
  {Holovko}}, \bibinfo {author} {\bibfnamefont {J.~P.}\ \bibnamefont
  {Badiali}}, \ and\ \bibinfo {author} {\bibfnamefont {D.}~\bibnamefont
  {di~Caprio}},\ }\href@noop {} {\bibfield  {journal} {\bibinfo  {journal} {J.
  Chem. Phys.}\ }\textbf {\bibinfo {volume} {123}},\ \bibinfo {pages} {234705}
  (\bibinfo {year} {2005})}\BibitemShut {NoStop}%
\bibitem [{\citenamefont {{Wheeler}}\ and\ \citenamefont
  {{Chandler}}(1971)}]{wheeler}%
  \BibitemOpen
  \bibfield  {author} {\bibinfo {author} {\bibfnamefont {J.}~\bibnamefont
  {{Wheeler}}}\ and\ \bibinfo {author} {\bibfnamefont {D.}~\bibnamefont
  {{Chandler}}},\ }\href@noop {} {\bibfield  {journal} {\bibinfo  {journal}
  {J.~Chem.~Phys.}\ }\textbf {\bibinfo {volume} {55}},\ \bibinfo {pages} {1645}
  (\bibinfo {year} {1971})}\BibitemShut {NoStop}%
\bibitem [{\citenamefont {Holovko}(2005)}]{holovko-soft}%
  \BibitemOpen
  \bibfield  {author} {\bibinfo {author} {\bibfnamefont {M.}~\bibnamefont
  {Holovko}},\ }in\ \href@noop {} {\emph {\bibinfo {booktitle} {Ionic Soft
  Matter: Modern Trends in Theory and Applications}}},\ \bibinfo {editor}
  {edited by\ \bibinfo {editor} {\bibfnamefont {D.}~\bibnamefont {Henderson}},
  \bibinfo {editor} {\bibfnamefont {M.}~\bibnamefont {Holovko}}, \ and\
  \bibinfo {editor} {\bibfnamefont {A.}~\bibnamefont {Trokhymchuk}}}\ (\bibinfo
   {publisher} {Springer},\ \bibinfo {address} {Berlin},\ \bibinfo {year}
  {2005})\ p.~\bibinfo {pages} {45}\BibitemShut {NoStop}%
\bibitem [{\citenamefont {{Di Caprio}}, \citenamefont {{Stafiej}},\ and\
  \citenamefont {{Badiali}}(2003)}]{di2003}%
  \BibitemOpen
  \bibfield  {author} {\bibinfo {author} {\bibfnamefont {D.}~\bibnamefont {{Di
  Caprio}}}, \bibinfo {author} {\bibfnamefont {J.}~\bibnamefont {{Stafiej}}}, \
  and\ \bibinfo {author} {\bibfnamefont {J.}~\bibnamefont {{Badiali}}},\
  }\href@noop {} {\bibfield  {journal} {\bibinfo  {journal} {Mol. Phys.}\
  }\textbf {\bibinfo {volume} {101}},\ \bibinfo {pages} {2545} (\bibinfo {year}
  {2003})}\BibitemShut {NoStop}%
\bibitem [{\citenamefont {{Di Caprio}}, \citenamefont {{Stafiej}},\ and\
  \citenamefont {{Badiali}}(1998)}]{di1998}%
  \BibitemOpen
  \bibfield  {author} {\bibinfo {author} {\bibfnamefont {D.}~\bibnamefont {{Di
  Caprio}}}, \bibinfo {author} {\bibfnamefont {J.}~\bibnamefont {{Stafiej}}}, \
  and\ \bibinfo {author} {\bibfnamefont {J.}~\bibnamefont {{Badiali}}},\
  }\href@noop {} {\bibfield  {journal} {\bibinfo  {journal} {J. Chem. Phys.}\
  }\textbf {\bibinfo {volume} {108}},\ \bibinfo {pages} {8572} (\bibinfo {year}
  {1998})}\BibitemShut {NoStop}%
\bibitem [{\citenamefont {{Kravtsiv}}\ \emph {et~al.}(2013)\citenamefont
  {{Kravtsiv}}, \citenamefont {{Holovko}}, \citenamefont {{Di Caprio}},\ and\
  \citenamefont {{Stafiej}}}]{preprint2013}%
  \BibitemOpen
  \bibfield  {author} {\bibinfo {author} {\bibfnamefont {I.}~\bibnamefont
  {{Kravtsiv}}}, \bibinfo {author} {\bibfnamefont {M.}~\bibnamefont
  {{Holovko}}}, \bibinfo {author} {\bibfnamefont {D.}~\bibnamefont {{Di
  Caprio}}}, \ and\ \bibinfo {author} {\bibfnamefont {J.}~\bibnamefont
  {{Stafiej}}},\ }\href@noop {} {\bibfield  {journal} {\bibinfo  {journal}
  {Preprint}\ }\textbf {\bibinfo {volume} {ICMP-13-01E}},\ \bibinfo {pages} {1}
  (\bibinfo {year} {2013})}\BibitemShut {NoStop}%
\bibitem [{\citenamefont {Hansen}\ and\ \citenamefont
  {McDonald}(2006)}]{hansen}%
  \BibitemOpen
  \bibfield  {author} {\bibinfo {author} {\bibfnamefont {{\relax J}.~{\relax
  P}.}\ \bibnamefont {Hansen}}\ and\ \bibinfo {author} {\bibfnamefont {{\relax
  I}.~{\relax R}.}\ \bibnamefont {McDonald}},\ }\href@noop {} {\emph {\bibinfo
  {title} {Theory of Simple Liquids}}}\ (\bibinfo  {publisher} {Academic
  Press},\ \bibinfo {address} {Oxford},\ \bibinfo {year} {2006})\BibitemShut
  {NoStop}%
\bibitem [{\citenamefont {Gahov}\ and\ \citenamefont {Cherski}(1978)}]{hakhov}%
  \BibitemOpen
  \bibfield  {author} {\bibinfo {author} {\bibfnamefont {{\relax
  F}.}~\bibnamefont {Gahov}}\ and\ \bibinfo {author} {\bibfnamefont {{\relax
  Y}.}~\bibnamefont {Cherski}},\ }\href@noop {} {\emph {\bibinfo {title}
  {Convolution-type equations}}}\ (\bibinfo  {publisher} {Nauka},\ \bibinfo
  {address} {Moscow},\ \bibinfo {year} {1978})\BibitemShut {NoStop}%
\bibitem [{\citenamefont {Amit}(1984)}]{Amit}%
  \BibitemOpen
  \bibfield  {author} {\bibinfo {author} {\bibfnamefont {{\relax
  D}.}~\bibnamefont {Amit}},\ }\href@noop {} {\emph {\bibinfo {title} {Field
  theory, the renormalization group, and critical phenomena}}}\ (\bibinfo
  {publisher} {World Scientific},\ \bibinfo {address} {Singapore},\ \bibinfo
  {year} {1984})\BibitemShut {NoStop}%
\bibitem [{\citenamefont {Zinn-Justin}(1989)}]{Zinn-Justin}%
  \BibitemOpen
  \bibfield  {author} {\bibinfo {author} {\bibfnamefont {{\relax
  J}.}~\bibnamefont {Zinn-Justin}},\ }\href@noop {} {\emph {\bibinfo {title}
  {Quantum Field Theory and Critical Phenomena}}}\ (\bibinfo  {publisher}
  {Clarendon Press},\ \bibinfo {address} {Oxford},\ \bibinfo {year}
  {1989})\BibitemShut {NoStop}%
\bibitem [{\citenamefont {Frenkel}\ and\ \citenamefont
  {Smit}(2002)}]{frenkel.book}%
  \BibitemOpen
  \bibfield  {author} {\bibinfo {author} {\bibfnamefont {{\relax
  D}.}~\bibnamefont {Frenkel}}\ and\ \bibinfo {author} {\bibfnamefont {{\relax
  B}.}~\bibnamefont {Smit}},\ }\href@noop {} {\emph {\bibinfo {title}
  {Understanding Molecular Simulations: From Algorithms to Applications}}}\
  (\bibinfo  {publisher} {Academic Press},\ \bibinfo {year} {2002})\BibitemShut
  {NoStop}%
\bibitem [{\citenamefont {Baker}\ and\ \citenamefont
  {Gravis-Morris}(1996)}]{baker}%
  \BibitemOpen
  \bibfield  {author} {\bibinfo {author} {\bibfnamefont {{\relax
  J}.}~\bibnamefont {Baker}}\ and\ \bibinfo {author} {\bibfnamefont {{\relax
  P}.}~\bibnamefont {Gravis-Morris}},\ }\href@noop {} {\emph {\bibinfo {title}
  {Pad\'{e} approximants}}}\ (\bibinfo  {publisher} {Cambridge U.P.},\ \bibinfo
  {year} {1996})\BibitemShut {NoStop}%
\end{thebibliography}%

\appendix

\section{The Riemann problem}\label{app:eigen}
Equation (\ref{eq}) can be represented in the form
\begin{align}
\label{Riemann}
P_{+}(K,\mu_1)\,{h}_{+}(K,\mu_1,\mu_2)-P_{-}(K,\mu_1)\,{h}_{-}(K,\mu_1,\mu_2)=
-L(\mu_2)\,\delta(\mu_1+\mu_2)\nonumber\\
\end{align}
where
\begin{eqnarray}
&&L(\mu_2)=4\pi\beta\left\{{A_1}{(\mu_2^2+p^2+\alpha_2^2)}
+{A_2}{(\mu_2^2+p^2+\alpha_1^2)}\right\},
\label{eq2.32}
\end{eqnarray}
\begin{eqnarray}
&&P_{+}(K,\mu_1)=(K^2+\mu_1^2+\alpha_1^2)(K^2+\mu_1^2+\alpha_2^2)+\nonumber\\
&&\qquad\qquad\qquad\qquad\varkappa_{1}^2
(K^2+\mu_1^2+\alpha_2^2)+\varkappa_{2}^2(K^2+\mu_1^2+\alpha_1^2),\qquad\label{eq2.33}\\
&&P_{-}(K,\mu_1)=(K^2+\mu_1^2+\alpha_1^2)(K^2+\mu_1^2+\alpha_2^2),\nonumber
\end{eqnarray}
Equation (\ref{Riemann}) is known as the Riemann problem
\cite{hakhov}. It can be solved by factorization, for which
purpose we write the fraction $P_{-}(K,\mu_1)/P_{+}(K,\mu_1)$
as
\begin{eqnarray}
\frac{P_{-}(K,\mu_1)}{P_{+}(K,\mu_1)}=\frac{Q_{+}(K,\mu_1)}{Q_{-}(K,\mu_1)},
\label{eq2.34}
\end{eqnarray}
where $Q_{+}(K,\mu_1),\,Q_{-}(K,\mu_1)$ are analytical
functions of $\mu_1$ and cannot be zero in the upper + or lower
- halves of the complex plane. The latter are easy to find:
\begin{eqnarray}
Q_{+}(K,\mu_1)=\frac{(\mu_1+i\alpha_{1}(K))(\mu_1+i\alpha_{2}(K))}
{(\mu_1+i\lambda_{2}(K))(\mu_1+i\lambda_{{1}}(K))},\nonumber\\
Q_{-}(K,\mu_1)=\frac{(\mu_1-i\lambda_{2}(K))(\mu_1-i\lambda_{{1}}(K))}
{(\mu_1-i\alpha_{1}(K))(\mu_1-i\alpha_{2}(K))},
\label{eq2.35}
\end{eqnarray}
where
\begin{eqnarray}
&&\alpha_{1}(K)=\sqrt{K^2+\alpha_{1}^2}\,,\qquad\alpha_{2}(K)=\sqrt{K^2+\alpha_{2}^2},\nonumber\\
&&\lambda_{2}(K)=\sqrt{K^2+\lambda_{2}^2}\,,\qquad\lambda_{{1}}(K)=\sqrt{K^2+\lambda_{{1}}^2}.
\label{eq2.36}
\end{eqnarray}
Coefficients $\lambda_{1},\,\,\lambda_{{2}}$ are found from
equation
\begin{eqnarray}
\lambda^4-(\alpha_1^2+\alpha_2^2+\varkappa_{1}^2+\varkappa_{2}^2)
\lambda^2+(\alpha_1^2+\varkappa_{1}^2)(\alpha_2^2+\varkappa_{2}^2)-
\varkappa_{1}^2\varkappa_{2}^2=0
\label{eq2.37}
\end{eqnarray}
giving
\begin{eqnarray}
\label{eq2.37.1}
\lambda_{1,2}^2=\frac{1}{2}\left(\varkappa_1^2+\alpha_1^2+\varkappa_2^2
+\alpha_2^2\pm\sqrt{\left(\varkappa_1^2+
\alpha_1^2-\varkappa_2^2-\alpha_2^2\right)^2+4\varkappa_1^2\varkappa_2^2}\right)
\end{eqnarray}
and coinciding with expressions (\ref{eigen}) obtained in the
framework of the mean field approximation.

We choose $i\lambda_{2}(K),\,\,i\lambda_{{1}}(K)$ to be in the
upper and $-i\lambda_{2}(K),\,\,-i\lambda_{{1}}(K)$ in the
lower halves of the analytical plane.

Equation (\ref{Riemann}) now reads
\begin{eqnarray}
\frac{{h}_{+}(K,\mu_1,\mu_2)}{Q_{+}(K,\mu_1)}-
\frac{{h}_{-}(K,\mu_1,\mu_2)}{Q_{-}(K,\mu_1)}\,=\,
-\,\frac{L(\mu_2)\,\delta(\mu_1+\mu_2)}{Q_{+}(K,-\mu_2)\,P_{+}(K,-\mu_2)}\,.
\label{eq2.38}
\end{eqnarray}
In~(\ref{Riemann}) the Dirac function is presented as the
difference of one-sided Dirac functions
\begin{eqnarray}
&&\delta(\mu_1+\mu_2)=\delta_{+}(\mu_1+\mu_2)-\delta_{-}(\mu_1+\mu_2),
\label{eq2.39}
\end{eqnarray}
which are analytical in the upper and lower halves of the
complex plane respectively. Since the index of~the problem
(\ref{Riemann}) is zero \cite{hakhov}, we obtain
\begin{eqnarray}
{h}_{+}(K,\mu_1,\mu_2)=
-\,\frac{L(\mu_2)Q_{+}(K,\mu_1)}{Q_{+}(K,-\mu_2)\,P_{+}(K,-\mu_2)}\delta_{+}(\mu_1+\mu_2)\nonumber\\
{h}_{-}(K,\mu_1,\mu_2)=
-\,\frac{L(\mu_2){Q_{-}(K,\mu_1)}}{Q_{+}(K,-\mu_2)\,P_{+}(K,-\mu_2)}\delta_{-}(\mu_1+\mu_2).
\label{eq2.444}
\end{eqnarray}
Replacing (\ref{eq2.32}), (\ref{eq2.34}) and (\ref{eq2.35}) into
(\ref{eq2.444}), we have
\begin{eqnarray}
&&{h}_{+}(K,\mu_1,\mu_2)\,=\nonumber\\
&&-4\pi\,\beta\frac{{A_1}{(\mu_2^2+\alpha_2^2)(K)}
+A_2{(\mu_2^2+\alpha_1^2(K))}}{(\mu_2-i\alpha_{1}(K))(\mu_2-i\alpha_{2}(K))
(\mu_2+i\lambda_{2}(K))(\mu_2+i\lambda_{{1}}(K))}\nonumber\\
&&\qquad\qquad\qquad\frac{(\mu_1+i\alpha_{1}(K))(\mu_1+i\alpha_{2}(K))}
{(\mu_1+i\lambda_{2}(K))(\mu_1+i\lambda_{{1}}(K))}\,\delta_{+}(\mu_1+\mu_2),
\label{eq2.4u!}
\end{eqnarray}
\begin{eqnarray}
&&{h}_{-}(K,\mu_1,\mu_2)\,=\nonumber\\
&&-4\pi\,\beta\frac{{A_1}{(\mu_2^2+\alpha_2^2)}
+A_2{(\mu_2^2+\alpha_1^2)}}
{(\mu_2-i\alpha_{1}(K))(\mu_2-i\alpha_{2}(K))
(\mu_2+i\lambda_{2}(K))(\mu_2+i\lambda_{{1}}(K))}\nonumber\\
&&\qquad\qquad\qquad
\frac{(\mu_1-i\lambda_{2}(K))(\mu_1-i\lambda_{{1}}(K))}
{(\mu_1-i\alpha_{1}(K))(\mu_1-i\alpha_{2}(K))}\,\delta_{-}(\mu_1+\mu_2).
\label{eq2.41}
\end{eqnarray}
Performing the inverse Fourier transformation
\begin{eqnarray}
&&h(R_{12},z_1,z_2)=
\int\frac{d\mathbf{K}}{(2\pi)^2}\mbox{\large{e}}^{-i\mathbf{K}\mathbf{R}_{12}}
\int\limits_{-\infty}^{\infty}\frac{d\mu_1}{2\pi}\mbox{\large{e}}^{-i\mu_1z_1}
\int\limits_{-\infty}^{\infty}\frac{d\mu_1}{2\pi}\mbox{\large{e}}^{-i\mu_2z_2}\nonumber\\
&&\qquad\qquad\qquad\qquad\qquad
\left\{{h}_{+}(K,\mu_1,\mu_2)-{h}_{-}(K,\mu_1,\mu_2)\right\},
\label{eq2.42}
\end{eqnarray}
we can find the originals of one-sided pair correlation
functions. Due to the considered model we are interested in the
case when both particles are in the upper half-space
$z_1>0,\,\,z_2>0$. We present one-sided $\delta$-functions as
\begin{eqnarray}
&&\delta_{+}(\mu_1+\mu_2)=\lim_{\varepsilon\rightarrow
0}\frac{i}{\mu_1+\mu_2+i\varepsilon}\,,\nonumber\\
&&\delta_{-}(\mu_1+\mu_2)=\lim_{\varepsilon\rightarrow
0}\frac{i}{\mu_1+\mu_2-i\varepsilon}\, \label{eq2.43}
\end{eqnarray}
and integrate by $\mu_1$. Then for $z_1>0$, closing the
integration contour in the lower half of the complex plane, we
have
\begin{eqnarray}
&&\lim_{\varepsilon\rightarrow 0}
\int\limits_{-\infty}^{\infty}\frac{d\mu_1}{2\pi}
\frac{(\mu_1+i\alpha_{1}(K))(\mu_1+i\alpha_{2}(K))}
{(\mu_1+i\lambda_{2}(K))(\mu_1+i\lambda_{{1}}(K))}\,
\frac{i}{\mu_1+\mu_2+i\varepsilon}\,\mbox{\large{e}}^{-i\mu_1z_1}\,=\nonumber\\
&&\frac{(\mu_2-i\alpha_{1}(K))(\mu_2-i\alpha_{2}(K))}
{(\mu_2-i\lambda_{2}(K))(\mu_2-i\lambda_{{1}}(K))}\,\mbox{\large{e}}^{i\mu_2z_1}-\nonumber\\
&&\qquad{i}\frac{(\lambda_{2}(K)-\alpha_{1}(K))(\lambda_{2}(K)-\alpha_{2}(K))}
{((\lambda_{2}(K)-\lambda_{{1}}(K))(\mu_2-i\lambda_{2}(K))}
\mbox{\large{e}}^{-\lambda_{2}(K)z_1}+\nonumber\\
&&\qquad\qquad{i}\frac{(\lambda_{{1}}(K)-\alpha_{1}(K))(\lambda_{{1}}(K)-\alpha_{2}(K))}
{(\lambda_{2}(K)-\lambda_{{1}}(K))(\mu_2-i\lambda_{{1}}(K))}
\mbox{\large{e}}^{-\lambda_{{1}}(K)z_1}. \label{eq2.44}
\end{eqnarray}
Now we integrate by $\mu_{2}$. We consider the case $z_{2}> 0$.
\begin{eqnarray}
&&
\int\limits_{-\infty}^{\infty}\frac{d\mu_2}{2\pi}
\frac{\left[{A_1}{(\mu_2^2+\alpha_2^2(K))}
+A_2{(\mu_2^2+\alpha_1^2(K))}\right]\mbox{\large{e}}^{-i\mu_2z_2}}
{(\mu_2-i\alpha_{1}(K))(\mu_2-i\alpha_{2}(K))
(\mu_2+i\lambda_{2}(K))(\mu_2+i\lambda_{{1}}(K))}\nonumber\\
&&\qquad\left\{\frac{(\mu_2-i\alpha_{1}(K))(\mu_2-i\alpha_{2}(K))}
{(\mu_2-i\lambda_{2}(K))(\mu_2-i\lambda_{{1}}(K))}\,\mbox{\large{e}}^{i\mu_2z_1}-\right.\nonumber\\
&&\left.\qquad\qquad{i}\frac{(\lambda_{2}(K)-\alpha_{1}(K))(\lambda_{2}(K)-\alpha_{2}(K))}
{(\lambda_{2}(K)-\lambda_{{1}}(K))(\mu_2-i\lambda_{2}(K))}
\mbox{\large{e}}^{-\lambda_{2}(K)z_1}+\right.\nonumber\\
&&\left.\qquad\qquad\qquad{i}\frac{(\lambda_{{1}}(K)-\alpha_{1}(K))(\lambda_{{1}}(K)-\alpha_{2}(K))}
{(\lambda_{2}(K)-\lambda_{{1}}(K))(\mu_2-i\lambda_{{1}}(K))}
\mbox{\large{e}}^{-\lambda_{{1}}(K)z_1}\right\}\,=\nonumber\\
&&-\frac{{A_1}{(\lambda_{2}^2-\alpha_2^2)}
+A_2{(\lambda_{2}^2-\alpha_1^2)}}{2\lambda_{2}(K)(\lambda_{{1}}^2-\lambda_{2}^2)}
\mbox{\large{e}}^{-\lambda_{2}(K)|z_1-z_2|}+\nonumber\\
&&\qquad\qquad\qquad\frac{{A_1}{(\lambda_{{1}}^2-\alpha_2^2)}
+A_2{(\lambda_{{1}}^2-\alpha_1^2)}}{2\lambda_{{1}}(K)(\lambda_{{1}}^2-\lambda_{2}^2)}
\mbox{\large{e}}^{-\lambda_{{1}}(K)|z_1-z_2|}+\label{eq2.45}\\
&&\frac{{A_1}{(\lambda_{2}^2-\alpha_2^2)}
+A_2{(\lambda_{2}^2-\alpha_1^2)}}{2\lambda_{2}(K)(\lambda_{{1}}(K)-\lambda_{2}(K))^2}\nonumber\\
&&\qquad\qquad\qquad\frac{(\lambda_{2}(K)-\alpha_1(K))(\lambda_{2}(K)-\alpha_2(K))}
{(\lambda_{2}(K)+\alpha_1(K))(\lambda_{2}(K)+\alpha_2(K))}
\mbox{\large{e}}^{-\lambda_{2}(K)(z_1+z_2)}-\nonumber\\
&&\frac{{A_1}{(\lambda_{{1}}^2-\alpha_2^2)}
+A_2{(\lambda_{{1}}^2-\alpha_1^2)}}
{(\lambda_{{1}}(K)+\lambda_{2}(K))(\lambda_{{1}}(K)-\lambda_{2}(K))^2}\nonumber\\
&&\qquad\qquad\qquad\frac{(\lambda_{2}(K)-\alpha_1(K))(\lambda_{2}(K)-\alpha_2(K))}
{(\lambda_{{1}}(K)+\alpha_1(K))(\lambda_{{1}}(K)+\alpha_2(K))}
\mbox{\large{e}}^{-\lambda_{2}(K)z_1-\lambda_{{1}}(K)z_2}-\nonumber\\
&&\frac{{A_1}{(\lambda_{2}^2-\alpha_2^2)}
+A_2{(\lambda_{2}^2-\alpha_1^2)}}
{(\lambda_{{1}}(K)+\lambda_{2}(K))(\lambda_{{1}}(K)-\lambda_{2}(K))^2}\nonumber\\
&&\qquad\qquad\qquad\frac{(\lambda_{{1}}(K)-\alpha_1(K))(\lambda_{{1}}(K)-\alpha_2(K))}
{(\lambda_{2}(K)+\alpha_1(K))(\lambda_{2}(K)+\alpha_2(K))}
\mbox{\large{e}}^{-\lambda_{{1}}(K)z_1-\lambda_{2}(K)z_2}+\nonumber\\
&&\frac{{A_1}{(\lambda_{{1}}^2-\alpha_2^2)}
+A_2{(\lambda_{{1}}^2-\alpha_1^2)}}{2\lambda_{{1}}(K)(\lambda_{{1}}(K)-\lambda_{2}(K))^2}\nonumber\\
&&\qquad\qquad\qquad\frac{(\lambda_{{1}}(K)-\alpha_1(K))(\lambda_{{1}}(K)-\alpha_2(K))}
{(\lambda_{{1}}(K)+\alpha_1(K))(\lambda_{{1}}(K)+\alpha_2(K))}
\mbox{\large{e}}^{-\lambda_{{1}}(K)(z_1+z_2)}\nonumber
\end{eqnarray}
Taking the inverse Fourier transform with respect to vector
$K$, we obtain the following expression for the case when
particles 1 and 2 are in the upper half-space, i.e. $z_{1}
> 0,\,z_{2} > 0$
\begin{eqnarray}
&&h_{+}(R_{12},z_1,z_2)=\beta
\frac{{A_1}{(\lambda_{2}^2-\alpha_2^2)}
+A_2{(\lambda_{2}^2-\alpha_1^2)}}{(\lambda_{{1}}^2-\lambda_{2}^2)}
\frac{\mbox{\large{e}}^{-\lambda_{2}R_{12}}}{R_{12}}-\\
&&\qquad\qquad\beta\frac{{A_1}{(\lambda_{{1}}^2-\alpha_2^2)}
+A_2{(\lambda_{{1}}^2-\alpha_1^2)}}{(\lambda_{{1}}^2-\lambda_{2}^2)}
\frac{\mbox{\large{e}}^{-\lambda_{{1}}R_{12}}}{R_{12}}-\nonumber\\
&&2\beta
\int\limits_{0}^{\infty}p\,J_{0}(KR_{12})\,dp \left\{\frac{{A_1}{(\lambda_{2}^2-\alpha_2^2)}
+A_2{(\lambda_{2}^2-\alpha_1^2)}}{2\lambda_{2}(K)(\lambda_{{1}}(K)-\lambda_{2}(K))^2}\right.\nonumber\\
&&\qquad\qquad\qquad\left.
\frac{(\lambda_{2}(K)-\alpha_1(K))(\lambda_{2}(K)-\alpha_2(K))}
{(\lambda_{2}(K)+\alpha_1(K))(\lambda_{2}(K)+\alpha_2(K))}
\mbox{\large{e}}^{-\lambda_{2}(K)(z_1+z_2)}-\right.\nonumber\\
&&\left.\frac{{A_1}{(\lambda_{{1}}^2-\alpha_2^2)}
+A_2{(\lambda_{{1}}^2-\alpha_1^2)}}
{(\lambda_{{1}}(K)+\lambda_{2}(K))(\lambda_{{1}}(K)-\lambda_{2}(K))^2}\right.\nonumber\\
&&\left.\qquad\qquad\qquad\frac{(\lambda_{2}(K)-\alpha_1(K))(\lambda_{2}(K)-\alpha_2(K))}
{(\lambda_{{1}}(K)+\alpha_1(K))(\lambda_{{1}}(K)+\alpha_2(K))}
\mbox{\large{e}}^{-\lambda_{2}(K)z_1-\lambda_{{1}}(K)z_2}-\right.\nonumber\\
&&\left.\frac{{A_1}{(\lambda_{2}^2-\alpha_2^2)}
+A_2{(\lambda_{2}^2-\alpha_1^2)}}
{(\lambda_{{1}}(K)+\lambda_{2}(K))(\lambda_{{1}}(K)-\lambda_{2}(K))^2}\right.\nonumber\\
&&\left.\qquad\qquad\qquad\frac{(\lambda_{{1}}(K)-\alpha_1(K))(\lambda_{{1}}(K)-\alpha_2(K))}
{(\lambda_{2}(K)+\alpha_1(K))(\lambda_{2}(K)+\alpha_2(K))}
\mbox{\large{e}}^{-\lambda_{{1}}(K)z_1-\lambda_{2}(K)z_2}\right.\nonumber\\
&&\left.
+\frac{{A_1}{(\lambda_{{1}}^2-\alpha_2^2)}
+A_2{(\lambda_{{1}}^2-\alpha_1^2)}}{2\lambda_{{1}}(K)(\lambda_{{1}}(K)-\lambda_{2}(K))^2}\right.\nonumber\\
&&\left.\qquad\qquad\qquad\frac{(\lambda_{{1}}(K)-\alpha_1(K))(\lambda_{{1}}(K)-\alpha_2(K))}
{(\lambda_{{1}}(K)+\alpha_1(K))(\lambda_{{1}}(K)+\alpha_2(K))}
\mbox{\large{e}}^{-\lambda_{{1}}(K)(z_1+z_2)}\right\},\nonumber
\label{umma}
\end{eqnarray}
where $J_{0}(KR_{12})$ is a Bessel function of the first kind.

\end{document}